\documentclass[usenatbib]{mn2e}
\usepackage{amsmath,graphics,amssymb}
\usepackage{subfigure}
\usepackage{times}

\newfont{\eufont}{eufm10}
\def\eu #1{\mbox{\eufont #1}}

\newcommand{\dmdt}{\mdot}

\newcommand{\vr}{{v_r}}
\newcommand{\pr}[1]{{#1}_{\mathrm{p}}}
\newcommand{\io}[1]{{#1}_{\mathrm{i}}}
\newcommand{\ha}[1]{{#1}_{h}}
\newcommand{\zav}[1]{\left(#1\right)}
\newcommand{\hzav}[1]{\left[#1\right]}
\newcommand{\xa}{\io{\eu Y}}
\newcommand{\kms}{\,\mathrm{km}\,\mathrm{s}^{-1}}

\newcommand{\Teff}{\mbox{$T_\mathrm{eff}$}}

\newcommand\mdot{\dot M}
\newcommand\smrok{\text{M}_{\odot}\,\mathrm{yr}^{-1}}
\newcommand\zm[1]{#1}

\DeclareMathAlphabet{\mathsc}{OT1}{cmr}{m}{sc}
\def\testbx{bx}%
\DeclareRobustCommand{\ion}[2]{%
\relax\ifmmode
\ifx\testbx\f@series
{\mathbf{#1\,\mathsc{#2}}}\else
{\mathrm{#1\,\mathsc{#2}}}\fi
\else{#1\,{\scshape{#2}}}%
\fi}

\title[NLTE models of line-driven stellar winds II.]
{NLTE models of line-driven stellar winds\\II. O stars in SMC}

\author[J.  Krti\v{c}ka]{Ji\v r\'\i\  Krti\v{c}ka\thanks{E-mail: krticka@physics.muni.cz}\\
\'Ustav teoretick\'e fyziky a astrofyziky P\v{r}F MU, CZ-611 37 Brno, Czech
Republic}

\begin{document}

\date{Received}

\maketitle

\begin{abstract}
We calculate NLTE line-driven wind models of selected O stars \zm{in the
spectral range of O4 to O9} in
the Small Magellanic Cloud (SMC). We compare predicted basic wind properties,
i.e. the terminal velocity and the mass-loss rate with values derived from
observation. We found relatively good agreement between theoretical and
observed terminal velocities. {On the other hand,} predicted mass-loss rates and mass-loss rates
derived from observation {are in} a good agreement only for higher mass-loss
rates. {Theoretical} mass-loss rates lower than approximately
$10^{-7}\,\text{M}_\odot\,\text{year}^{-1}$ are significantly higher than those
derived from observation. {These results confirm the previously reported
problem of weak winds, since our calculated mass-loss rates are in a fair
agreement with predictions of Vink et al.~(2001).} 
We study multicomponent models for these winds.
For this purpose we develop a more detailed {description of}
wind decoupling. {We show that the instability connected with} the decoupling of individual wind
elements {may occur for low-density winds}. 
In the case of winds with {very} low observed mass-loss rate the multicomponent
effects are important for the wind structure, however this is not able to
consistently explain the difference between predicted mass-loss rate and
mass-loss rate derived from observation {for these stars}.
{Similar to previous studies, we found the
dependence of wind parameters on the metallicity.} We conclude that the wind
mass-loss rate 
significantly {increases with} metallicity {as $\dmdt\sim Z^{0.67}$}, 
whereas the wind terminal velocity on
average depends on metallicity only slightly, {namely $v_\infty\sim
Z^{0.06}$} (for studied {stars}).
\end{abstract}



\section{Introduction}

In recent few years 8m class telescopes became routinely available for {the}
stellar
research. This enabled {detailed} study of many stars in the Local Group. Clearly, many
astrophysically important types of stars can be studied from another
perspective. This is especially true for stars from the Small Magellanic
Cloud (SMC). Their lower metallicity compared to the solar value
({the average metallicity of individual elements is} $Z/Z_\odot=0.2$, e.g.
Venn \citeyear{ven}, Bouret at al. \citeyear{bourak}) enables {to}
study {the} stellar properties and evolution with respect to metallicity. 

One of the most important properties of hot stars, that can influence both
{the} observed spectrum and {the} stellar evolution, is the presence of the stellar wind.
The existence of significant dependence of the basic
wind properties (i.e. {the} {amount of mass expelled from the star per unit of
time, the} mass-loss rate and the wind velocity \zm{at large distances from the
star}, {the} terminal velocity) on the metallicity
has been anticipated already at the very beginning of the theoretical study of
hot star winds {(Abbott \citeyear{abpar}, Kudritzki et al. \citeyear{kupapu})}.
Although the most abundant elements in the observed universe,
hydrogen and helium mostly contribute to the wind density, their contribution to
the radiative acceleration is very small (e.g. Abbott \citeyear{abpar}).
Heavier elements like carbon, nitrogen, oxygen or iron {are much more
important for the wind acceleration \zm{of present hot stars}. This is because heavier elements}
have {effectively} much more lines that are available to absorb the stellar
radiation and to accelerate the stellar wind. 
Clearly, in most cases the radiative force shall be
higher for higher metallicity.
Consequently, the basic wind properties shall depend on {the} metallicity.

Although relatively simple expressions for the dependence of the mass-loss rate
and {the} terminal velocity on the metallicity can be obtained even on the basis of
the standard CAK (Castor, Abbott \& Klein \citeyear{cak}) theory (e.g. Puls et al.
\citeyear{puspo}), the detailed dependence is probably more complicated (Puls et al.
\citeyear{puspo}, Vink et al. \citeyear{vikolamet}). It is clear that more precise 
(NLTE) wind models are necessary to study this problem in detail. 

{There are several independent NLTE wind models that can be used for the
study of the metallicity dependence of the wind properties. These codes differ
in the level of sophistication of the treatment of the wind problem. One of the
most important aspects that influences the reliability of individual  NLTE
wind models is
the inclusion of proper bound-free and bound-bound blanketing in the UV domain.
While models of Vink et al. (\citeyear{vikolamet},  hereafter VKL) use Monte Carlo method for
the
solution of the radiative transfer equation, models of  Pauldrach et al.
(\citeyear{pahole}) solve this equations in detail. Wind models of
Gr\"afener \& Hamann (\citeyear{graham}), that were successfully used for the
calculation of wind models of WR stars, use comoving-frame formulation of the radiative
transfer equation. All these models to some extent properly take into account
blanketing effects in the UV domain that are important for the
correct calculation of the ionization balance and of the
radiative force. Hence, these models are able to reliably predict the
most important wind parameter -- the  mass-loss rate and its metallicity
dependence.

However, also the wind velocity field and especially the terminal velocity
can be influenced by the metallicity. To study this effect, the solution of
hydrodynamic equations is necessary. Since models of VKL, that are widely used
in hot star evolutionary calculations, do not solve these
equations, these models assume a prespecified velocity law and
cannot be used for the predictions of the velocity structure. This is an
important difference between models of VKL and e.g.~Pauldrach et al.
(\citeyear{pahole}) or Gr\"afener \& Hamann
(\citeyear{graham}). Hydrodynamical
wind models for different metallicities were calculated by Kudritzki
(\citeyear{kudmet}). According to these models the wind terminal velocity
decreases with decreasing metallicity, in agreement with SMC observations (see
an extensive compilation of Kudritzki \& Puls \citeyear{kupul}).

However, the wind metallicity does not vary the radiative force only. The
momentum acquired by the heavier elements from the radiative field is
transferred to the bulk wind component (i.e.~hydrogen and helium) via the
Coulomb collisions. Since the frictional force caused by these collisions
depends on number densities of heavier elements and hydrogen-helium component,
the effects induced by the collisions are more important
in the low-metallicity environment (Kudritzki \citeyear{kudmet}, Krti\v cka et al.
\citeyear{gla}). NLTE models available to study this problem were
presented by  Krti\v{c}ka \& Kub\' at (\citeyear{nltei}, hereafter KK1).

Some hot stars with low luminosities (and consequently low mass loss rates)
show signatures of very weak winds, much
weaker than that deduced from standard wind theory (Bouret et al.
\citeyear{bourak}; Martins et al. \citeyear{martin,okali}). Since this
discrepancy is based mostly on wind models of Vink et al.
(\citeyear{vikolamet}), it would be interesting to \zm{test} this result also with
independent NLTE wind models and to test whether this discrepancy can be
caused by multicomponent effects.}

{T}he study of the metallicity dependence of stellar wind properties is
not important only for our understanding of stellar winds or {massive} stars itself.
Since stellar winds
play a role in the chemical enrichment of galaxies and are responsible for input
of a large amount of momentum and energy in the interstellar medium, the
detailed knowledge of variations of basic wind parameters with metallicity is
important also for other fields of astrophysics.

As we have discussed, wind properties of hot stars shall depend on the
stellar metallicity. {L}ow-metallicity environment of SMC offers a unique
possibility to test these available predictions.  
To do so, we present here wind models
for selected SMC hot stars.

\section{Model description}

\begin{table}
\caption{Atoms and {ions} included in the NLTE calculations.
In this table, level means either an individual level or a
set of levels merged into a superlevel.}
\label{prvky}
\centering
\begin{tabular}{l@{\hspace{-0.7mm}}r@{\hspace{7mm}}l@{\hspace{-0.7mm}}r@{\hspace{7mm}}l@{\hspace{-0.7mm}}r}
\hline\hline
Ion & Levels & Ion & Levels & Ion & Levels \\
\hline
 \ion{H}{i}   &   9  &  \ion{Ne}{iv} &  12 &  \ion{S}{iv}   &  18  \\
 \ion{H}{ii}  &   1  &  \ion{Ne}{v}  &  17 &  \ion{S}{v}    &  14  \\
 \ion{He}{i}  &  14  &  \ion{Ne}{vi} &   1 &  \ion{S}{vi}   &  16  \\
 \ion{He}{ii} &  14  &  \ion{Na}{ii} &  13 &  \ion{S}{vii}  &   1  \\
 \ion{He}{iii}&   1  &  \ion{Na}{iii}&  14 &  \ion{Ar}{iii} &  25  \\
 \ion{C}{ii}  &  14  &  \ion{Na}{iv} &  18 &  \ion{Ar}{iv}  &  19  \\
 \ion{C}{iii} &  23  &  \ion{Na}{v}  &  16 &  \ion{Ar}{v}   &  16  \\
 \ion{C}{iv}  &  25  &  \ion{Na}{vi} &   1 &  \ion{Ar}{vi}  &   1  \\
 \ion{C}{v}   &   1  &  \ion{Mg}{ii} &  14 &  \ion{Ca}{ii}  &  16  \\
 \ion{N}{ii}  &  14  &  \ion{Mg}{iii}&  14 &  \ion{Ca}{iii} &  14  \\
 \ion{N}{iii} &  32  &  \ion{Mg}{iv} &  14 &  \ion{Ca}{iv}  &  20  \\
 \ion{N}{iv}  &  23  &  \ion{Mg}{v}  &  13 &  \ion{Ca}{v}   &  22  \\
 \ion{N}{v}   &  13  &  \ion{Mg}{vi} &   1 &  \ion{Ca}{vi}  &   1  \\
 \ion{N}{vi}  &   1  &  \ion{Al}{iii}&  14 &  \ion{Fe}{iii} &  29  \\
 \ion{O}{ii}  &  50  &  \ion{Al}{iv} &  14 &  \ion{Fe}{iv}  &  32  \\
 \ion{O}{iii} &  29  &  \ion{Al}{v}  &   1 &  \ion{Fe}{v}   &  30  \\
 \ion{O}{iv}  &  39  &  \ion{Si}{iii}&  12 &  \ion{Fe}{vi}  &  27  \\
 \ion{O}{v}   &  14  &  \ion{Si}{iv} &  13 &  \ion{Fe}{vii} &   1  \\
 \ion{O}{vi}  &  20  &  \ion{Si}{v}  &  15 &  \ion{Ni}{iii} &  36  \\
 \ion{O}{vii} &   1  &  \ion{Si}{vi} &   1 &  \ion{Ni}{iv}  &  38  \\
 \ion{Ne}{ii} &  15  &  \ion{S}{ii}  &  14 &  \ion{Ni}{v}   &  48  \\
 \ion{Ne}{iii}&  14  &  \ion{S}{iii} &  10 &  \ion{Ni}{vi}  &   1  \\
\hline
\end{tabular}
\end{table}

The models applied in this paper were in detail described by 
KK1. Here we only summarise the basic model
properties and refer an interested reader to this paper.

{M}odels assume spherically symmetric and 
{stationary} stellar wind. Occupation
numbers of selected {atoms and} ions {(see Table~\ref{prvky})}
are obtained by the solution of statistical equilibrium
(NLTE) equations together with the radiative transfer equation.
{The radiative transfer equation in lines is solved in the Sobolev
approximation (Sobolev \citeyear{sobolevprvni}, Castor \citeyear{cassob}),
whereas the continuum  radiative transfer equation is solved by
Feautrier method in
the spherical coordinates (see Mihalas \& Hummer
\citeyear{sphermod} or Kub\'at \citeyear{dis}).}
Derived occupation numbers are used to calculate the radiative
force (in the Sobolev approximation) and the radiative cooling/heating term
(using the thermal balance of electrons
method, Kub\'at et al. (\citeyear{kpp})). This enables to 
solve the hydrodynamic equations. These procedures are iterated to obtain
consistent model structure. Finally, wind mass-loss rates and terminal
velocities of studied stars can be obtained from our models and compared with
observations.

Although our models involve some simplifying assumptions (especially the
simplified treatment of the radiative transfer \zm{(splitting of the radiative
transfer in continuum and in the lines, neglect of line overlaps)}
compared to more advanced
models of e.g. Pauldrach et al. (\citeyear{pahole}), {VKL or Gr\"afener \&
Hamann (\citeyear{graham})),}
they were able to correctly predict basic wind parameters
of late O stars (see KK1) and of A supergiant (see Krti\v{c}ka \&
Kub\' at \citeyear{velebobr}). Moreover, our models
have some advantages compared to {some} other models available in the literature, for
example the direct calculation of the radiative force without using of force
multipliers or multicomponent treatment of model equations (see Krti\v cka \&
Kub\'at \citeyear{kkii}, hereafter KKII).

Presented models were only slightly modified with respect to the status
described by KK1. First, we included accelerated lambda iterations in
continuum (or, more precisely, approximate Newton-Raphson iterations) based on
Rybicki \& Hummer (\citeyear{rybashumremali}) paper and Ng acceleration (Ng
\citeyear{ng}) to accelerate the convergence of
the system of statistical equilibrium
equations (see also Hillier \&  Miller (\citeyear{hilmi}), or Hubeny
(\citeyear{hubtueb}) for a review). This will be described in
a following paper (Krti\v cka \& Kub\'at, in preparation).
Second, our previous set
of {included} atomic models based on TLUSTY files (Hubeny \citeyear{tlusty}, Hubeny
\& Lanz \citeyear{hublaj}, Hubeny \& Lanz
\citeyear{hublad}, Lanz \& Hubeny \citeyear{lahub}) or Opacity Project (Seaton
\citeyear{top}, Luo \& Pradhan \citeyear{top1}, Sawey \& Berrington
\citeyear{savej}, Seaton et al. \citeyear{topt}, Butler et al. \citeyear{bumez},
Nahar \& Pradhan \citeyear{napra}) and Iron Project (Hummer et al.
\citeyear{zel0}, Bautista \citeyear{zel6}, Nahar \& Pradhan \citeyear{zel2},
Zhang \citeyear{zel1}, Bautista \& Pradhan \citeyear{zel5}, Zhang \& Pradhan
\citeyear{zel4}, Chen \& Pradhan \citeyear{zel3}) data was slightly extended
({see} Table~\ref{prvky}). {T}hese 
mentioned changes do not significant{ly} influence the derived
results.
{T}o complete the list of atomic databases used,  {the}
oscillator strengths necessary for the calculation of the radiative
force are extracted from the VALD database (Piskunov et al. 
\citeyear{vald1}, Kupka et al.  \citeyear{vald2}).

{T}he boundary radiative flux is taken from grid of line-blanketed 
plane-parallel model atmospheres OSTAR2002 (Lanz \&
Hubeny \citeyear{lahub}) instead of {from} H-He spherically symmetric models of Kub\'at
(\citeyear{kub}).  Since line-blanketed fluxes have generally lower flux in the UV
region (where are many lines important for radiative driving), obtained
mass-loss rates are slightly lower {(roughly $1.4\times$)}
than that derived using H-He fluxes.
However, this difference is lower than the dispersion of
mass-loss rates for Galactic O stars (see KK1)  and does not significantly
influence predicted wind terminal velocit{ies} {(the difference between the
terminal velocities is about $100\kms$)}.


\section{Calculated wind models}

\subsection{Parameters of studied SMC star{s}}

{Since our main intention for future studies is to study weak winds of B
stars, for the present study we}
selected only those {cooler (i.e.~with effective temperatures
$\Teff\lesssim42\,000\,$K)} O SMC stars, for which at least
reliable {estimate} of their mass-loss rate is available in the literature.
{We tried to omit those stars for which their wind parameters are
uncertain and which are binaries. Moreover, we also aim to base the list on
broader surveys with larger number of individual stars studied.}
Stellar parameters and wind parameters of these stars are given in
Tab.~\ref{mmmhvezpar}. Stellar effective temperatures and radii are taken from
Puls et al. (\citeyear{pulmoc}, hereafter P96), Bouret et al. (\citeyear{bourak},
hereafter B03)
and Massey et al. (\citeyear{maso}, hereafter M04). Whereas parameters given by P96 were
obtained by models without wind-blanketing, parameters given by B03 and M04
were derived using models with wind-blanketing. {To study the thin-wind
problem in detail we added also stars from Martins et al. (\citeyear{martin},
hereafter Mr04) sample. \zm{These stars are suspected Vz stars,
i.e.~stars close to the ZAMS.} \zm{They} exhibit much lower wind
spectral signature than that predicted from standard theory. On the other hand
there are only upper limits of their observed \zm{mass-loss rates and lower
limits of their observed terminal velocities available.}

{SMC stars are in some sense more suitable for the test of theoretical models
than stars from our Galaxy. Since the distance to SMC is known with relatively
high precission, the stellar radius and mass-loss rate may be also derived more
reliably.}

For our study we adopted evolutionary stellar masses either derived by B03 or by
us using evolutionary tracks of stars with initial metallicity $Z/Z_\odot=0.2$
calculated by Charbonnel et al. (\citeyear{uhlak}). {The use of the
evolutionary masses is a relatively important assumption that can significantly
influence the results derived due to the discrepancy between stellar masses of
hot stars derived from evolutionary tracks and from spectroscopy (Herrero et al.
\citeyear{hekuku}). Since KK1 in their analysis used evolutionary masses, we also
use evolutionary masses here, but in fact
for many stars from KK1 sample these masses are nearly equal. Our SMC sample is not
homogeneous in that sense, since for some stars these masses are equal, but for
a significant part of stars the evolutionary mass is roughly $1.5\times$ higher than the
spectroscopic one. The use of spectroscopic masses instead of the evolutionary ones
would help to obtain a better agreement between observation and theory for some
stars, however it would cause differences for some others.}

When available, the {abundances}
of individual {elements were taken} from the literature, however in other cases we
{assumed} average value $Z/Z_\odot=0.2$ derived for SMC stars
(e.g. Venn \citeyear{ven}, M04). {We use Galactic helium abundance.}

{There are indications that stellar winds of O stars are clumped
(e.g.~B03). From the observational point of view, the possible wind clumping decreases
the wind mass-loss rate inferred from the observation because the line profiles of
clumped wind mimic those with higher mass-loss rate. 
According to the numerical simulations of wind instability (e.g.~Feldmeier
et al. \citeyear{felpulpal}, Runacres  \& Owocki \citeyear{runow}) the
theoretical mean mass-loss rate is nearly the same for smooth and structured winds.
Thus, if the observations really show signatures of clumping at all spectral
regions where the stellar wind is observed, then, according to our present
knowledge, the theoretically predicted values
of mass-loss rates should basically correspond to the values derived
from observation with account of clumping. For our study we adopted the values
derived from observations assuming "smooth" winds partly because these values
were for larger SMC sample (to our knowledge) derived only by B03 and partly because
the models of wind clumping are still schematic.}

\begin{table*}
\centering
\caption{Stellar and wind parameters of selected SMC stars.
Stellar parameters (radius $R_{*}$, {the} effective temperature
$\Teff$ and {the} metallicity relative to {the} solar {value} $Z/Z_\odot$) 
were adopted either from Puls et al. (\protect\citeyear{pulmoc}, hereafter P96),
Bouret et al. (\protect\citeyear{bourak}, hereafter B03), Massey et al. 
(\protect\citeyear{maso}, hereafter M04) {and Martins et al.
(\citeyear{martin},   hereafter Mr04)}. 
{For the stellar mass $M$ we assume values derived using evolutionary
tracks.}
For stars with metallicities denoted as B03 {and Mr04}
we adopted detailed abundance determinations from B03 {and Mr04} for C, N, O, Si, S and Fe
and $Z/Z_\odot=0.2$ for other {heavier} elements.  Observed wind parameters (i.e.
the mass-loss
rates $\mdot$ and the wind terminal velocities $v_{\infty}$) were also mostly taken
from P96, B03, M04 {and Mr04} (when available), however see the discussion for individual
stars. Predicted values of wind parameters were derived by our code.
For stars for which $v_\infty$ value is {not given in the table} authors
provide only lower limits {that is in agreement with o}ur predicted value of $v_\infty$.}
\label{mmmhvezpar}
\begin{tabular}{llcccccccccc}
\hline
\hline
\multicolumn{1}{c}{Star} & \multicolumn{5}{c}{Stellar parameters}  &
\multicolumn{2}{c}{Mass loss rates $\mdot$} &
\multicolumn{2}{c}{Terminal velocities $v_{\infty}$} & Source \\
& {Sp.} & $R_{*}$ & $M$ & $\Teff$ & $Z/Z_\odot$ &
observed & predicted &
observed  & 
predicted\\
 &&$[\text{R}_{\odot}]$ &  $[\text{M}_{\odot}]$ & $[\mathrm{K}] $  & &
$[\smrok]$ &
$[\smrok]$ &
$[\kms]$ & 
$[\kms]$ \\
\hline
{NGC 346 WB 1} &  {O4III} & $23.3$ & $95$ & $42\,000$ & $0.2$ &
  $4.8\,10^{-6}$ & $3.5\,10^{-6}$ & $2250$ & $2240$ & P96\\
{NGC 346 WB 4} &  {O5.5V} & $14.2$ & $53$ & $42\,000$ & $0.2$ &
  $\leq1\,10^{-7}$ & $7.8\,10^{-7}$ & $1950$ & $2200$ & P96\\
{NGC 346 WB 6} &  {O4V} & $11.2$ & $40$ & $41\,500$ & B03 &
  $2.7\,10^{-7}$ & $3.5\,10^{-7}$ & $2300$ & $2180$ & B03\\
{NGC 346 MPG 368} &  {O5.5V}& $10.6$ & $38$ & $40\,000$ & B03 &
  $1.5\,10^{-7}$ & $1.8\,10^{-7}$ & $2100$ & $2580$ & B03\\
{NGC 346 MPG 113} &  {O6V} & $7.8$ & $33$ & $40\,000$ & B03 &
  $3\,10^{-9}$ & $4.8\,10^{-8}$ & & $3370$  & B03\\
{N81 \#2}&O6.5&$7.9$ &$31$& $40\,000$ &Mr04 &
  $\lesssim10^{-8}$ & $4\,10^{-8}$ & & $2880$ & Mr04\\
{AzV 75} &  {O5III} & $25.4$ & $92$ & $40\,000$ & $0.2$ &
  $3.5\,10^{-6}$ & $3.8\,10^{-6}$ & $2100$ & $2120$ & M04\\
{N81 \#1}&O7  &$10.3$&$34$& $38\,500$ & Mr04 &
  $\lesssim10^{-8}$ & $9\,10^{-8}$ & & $2530$ & Mr04\\
{AzV 26} &  {O6I} & $27.5$ & $86$ & $38\,000$ & $0.2$ &
  $2.5\,10^{-6}$ & $3.8\,10^{-6}$ & $2150$ & $1850$ & M04 \\
{AzV 232} &  {O7Ia} & $29.3$ & $93$ & $37\,500$ & $0.2$ &
  $5.5\,10^{-6}$ & $4.1\,10^{-6}$ & $1400$ & $1880$  & P96\\
{AzV 207} &  {O7V} & $11.0$ & $33$ & $37\,000$ & $0.2$ &
  $1\,10^{-7}$ & $1.0\,10^{-7}$ & $2000$ & $2060$ & M04 \\
{N81 \#11}&O7.5&$6.9$&$23$& $37\,000$ & Mr04 &
 $\lesssim10^{-9}$ & $2\,10^{-8}$ & & $2270$ & Mr04 \\
{N81 \#3}&O8.5&$5.0$ &$19$& $36\,000$ & Mr04 &
 $\lesssim3\,10^{-9}$ & $3\,10^{-9}$ & & $2260$& Mr04\\
{AzV 238} &  {O9III} & $15.5$ & $37$ & $35\,000$ & $0.2$ &
  $1.3\,10^{-7}$ & $2.8\,10^{-7}$ & $1200$ & $1830$  & P96\\
{NGC 346 MPG 487} & {O6.5V} & $10.2$ & $25$ & $35\,000$ &B03 &
  $3\,10^{-9}$ & $3.0\,10^{-8}$ & & $2300$ &B03 \\
{AzV 469} &  {O8.5II} & $21.2$ & $38$ & $32\,000$ & $0.2$ &
  $1.8\,10^{-6}$ & $4.0\,10^{-7}$ & $2000$ & $2010$ & M04 \\
{NGC 346 MPG 12} &  {O9.7V}&  $10.1$  & $21$ & $31\,000$ &B03 &
  $1\,10^{-10}$ & $1.7\,10^{-8}$ & & $1820$ &B03 \\
\hline
\end{tabular}
\end{table*}

Wind parameters adopted for the comparison with theoretical values will be
discussed individually for those star{s}, for which it is necessary.
{Note that in the text we will term the mass-loss rate estimated from
observation as "observed mass-loss rate", although one should keep in mind that
this quantity cannot be directly derived from spectra and that it is
model-dependent.}

\paragraph*{NGC 346 WB 1} 
This is a multiple system (Heydari-Malayeri \& Hutsem\'ekers \citeyear{hemalhut}).
The terminal velocity derived for this system by {P96}
$v_\infty=2650\kms$ is marked as uncertain due to the complexity of {the} absorption
profile. {P96} note that another possible value of {the} terminal velocity is
$v_\infty=2250\kms$. Prinja \& Crowther (\citeyear{rampa})
obtained value of edge velocity (velocity at which the line profile meets the
continuum level) as $v_\mathrm{edge}=2830\kms$ for \ion{N}{v} lines. Using their
derived approximate relation $v_\infty=0.8v_\mathrm{edge}$ we obtain terminal
velocity $v_\infty=2260\kms$. Thus, we adopted $v_\infty=2250\kms$ {as a}
value of {the} terminal velocity.

\paragraph*{NGC 346 WB 4}
The observed terminal velocity by {P96}
$v_\infty=1550\kms$ is quoted as
uncertain. The typical terminal velocity for stars of similar 
spectral type is higher, the terminal velocity calculated as twice the escape
velocity (roughly suitable for SMC stars, B03) is $1900\kms$. Using approximate
relation $v_\infty=0.8v_\mathrm{edge}$ of Prinja \& Crowther (\citeyear{rampa})
and their measurement of edge velocity of \ion{C}{iv} lines we obtain
$v_\infty=1950\kms$. Hence, we adopt this value of terminal velocity.

\paragraph*{NGC 346 WB 6}
The edge velocity obtained for this star by Prinja \& Crowther (\citeyear{rampa})
$v_\mathrm{edge}=1925\kms$ is lower than the terminal velocity
$v_\infty=2250\kms$
derived by P96 or $v_\infty=2300\kms$ by B03.
This difference probably illustrates the fact 
that the correct determination of wind velocities for SMC is difficult due to
their low wind density.

This star was independently studied by P96, who
obtained slightly lower value of {the} effective temperature than B03
($\Teff=40\,000\,$K). 

\paragraph*{AzV 232}
Crowther et al. (\citeyear{studeny}) applied models with wind blanketing to study
{the} 
stellar parameters of this star and obtained much lower value of the effective
temperature than P96 ($\Teff=32\,000\,\text{K}$). However, we calculate{d}
wind parameters with this new determination of stellar parameters (i.e. the
effective temperature, stellar mass, radius and abundances of C, N, O) {and}
we obtain{ed}
too low value of the mass loss rate ($\mdot=9\,10^{-7}\,\smrok$), much lower 
than the observed value. Also models of VKL
predict with this new parameters about five times lower mass-loss rate than
the observed value. Probably, this may be caused by the fact that this star 
has {a} very peculiar chemical composition ($Z(\text{C})/Z_\odot(\text{C})=0.07$,
$Z(\text{O})/Z_\odot(\text{O})=0.1$, whereas $Z(\text{N})/Z_\odot(\text{N})=2.0$,
Crowther et al. \citeyear{studeny}) and abundances of other elements which are also
important for driving of the stellar wind were not determined. {Thus, to keep
our sample more homogeneous, we used the stellar parameters derived by P96.}


\subsection{Comparison of calculated wind parameters with observation}
\label{srovnavacka}

\begin{figure}
\begin{center}
\resizebox{\hsize}{!}{\includegraphics{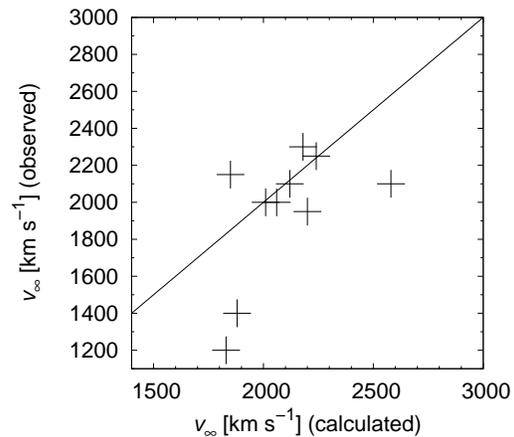}}
\end{center}
\caption[]{Comparison of calculated and observed 
terminal velocities. {Stars for which there is no reliable estimate of their
terminal velocities (in fact those stars that exhibit very weak winds)
were excluded from the plot.}
Line denotes one to one relation.}
\label{vnek}
\end{figure}

\begin{figure}
\resizebox{\hsize}{!}{\includegraphics{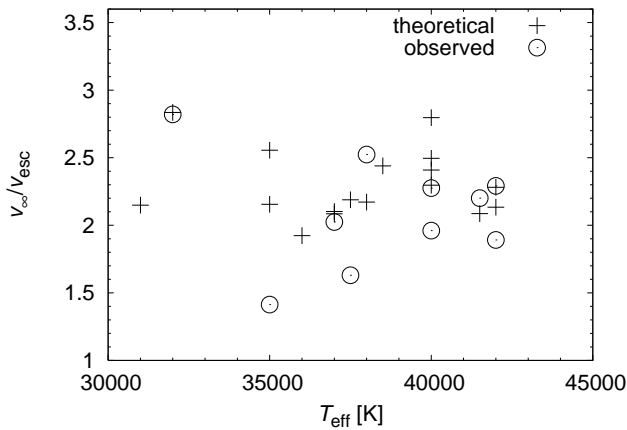}}
\caption[]{Comparison of the ratio of the wind terminal velocity $v_\infty$
to the surface escape velocity $v_\text{esc}$ derived using theoretical and 
observed values of $v_\infty$. {Note that the observed values for stars with
only lower limit of the terminal velocity available (missing $v_\infty$  
in Tab.~\ref{mmmhvezpar}) are not plotted in this graph.}}
\label{vnekvuni}
\end{figure}

\begin{figure}
\resizebox{\hsize}{!}{\includegraphics{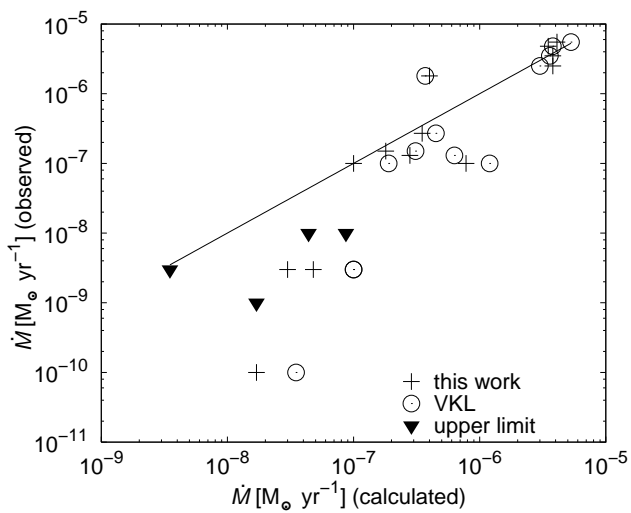}}
\caption[]{Comparison of calculated mass-loss rates {(either by us or by
VKL)} and mass-loss rates derived
from observation. {Points corresponding to the upper limits were obtained
using our model predictions and upper limits derived from observations.}
Line denotes one to one relation.}
\label{dmdt} 
\end{figure}


In Fig.~\ref{vnek} we compare calculated wind terminal velocities for selected
stars with observed values. {W}hile for some stars there is a very good
agreement between observed and calculated {terminal velocities}, for some stars the
agreement is {worse}. Before we will discuss this problem in a next section
let us conclude, that derived scatter {between} observed and calculated wind terminal
velocities is {slightly} higher than that for Galactic stars (see KK1).

{Generally, it is difficult to assess the accuracy of determination of 
wind parameters. P96 state that due to the edge variability and
contamination of the edges of wind absorption lines by underlying photospheric
lines the terminal velocities should be regarded as accurate up to $\pm10\,\%$.
Taking this as a typical error of terminal velocity determination we conclude,
that with exception of stars NGC~346~MPG 368, 
AzV~26, AzV~232 and AzV~238 the rest of terminal
velocity estimation{s} lie within the mentioned uncertainty interval.}

According to the predictions of stellar wind theory the wind terminal velocity
$v_\infty$ is correlated with {the} surface escape velocity $v_\text{esc}$.
{For Galactic O stars the value of $v_\infty/v_\text{esc}$ does not depend
on {their effective} temperature and has a mean value of about $2.5$ (Lamers et al. \citeyear{lsl}).}
Thus, we also plot the ratio of $v_\infty/v_\text{esc}$ for studied stars (see
Fig.~\ref{vnekvuni} where we compare ratio of $v_\infty/v_\text{esc}$ derived
using theoretical and observed values of $v_\infty$). 
The mean value of $v_\infty/v_\text{esc}\sim2.3$ is slightly
lower than that obtained by KK1 for Galactic stars. 
Our calculated $v_\infty/v_\text{esc}$ ratio is
slightly lower than recent observational finding of Evans et al. (\citeyear{eva}),
who derived the median $v_\infty/v_\text{esc}=2.63$ for SMC stars 
with effective temperature{s} higher than $24\,000\,\text{K}$. Since B03
obtained for their (however limited) sample $v_\infty/v_\text{esc}\sim2.3$
we conclude that our {calculations} {are} in agreement with other studies
{and that our results may indicate that SMC {hot stars} terminal velocities are slightly
lower than that of Galactic {hot} stars}.

{Similarly to Galactic O stars,}
there is {a} large scatter of {both observed and theoretical}
$v_\infty/v_\text{esc}$ values {(see Fig.~\ref{vnekvuni}).}
The scatter of
theoretical values of $v_\infty/v_\text{esc}$ is {slightly}
higher {for SMC stars} than {for} Galactic
stars (KK1). {The origin of this scatter is probably {partly}
connected with high sensitivity
of terminal velocities on detailed wind parameters in the outer wind regions
(see Puls et al. \citeyear{pusle} and also Sect.~\ref{citlivka})} {and also
with uncertain values of the terminal velocities.}

The comparison of calculated mass-loss rates and mass-loss
rates derived from observations in Fig.~\ref{dmdt} shows relatively good
agreement for stars with higher mass-loss rates ($\dmdt
\gtrsim10^{-7}\,\smrok)$, in many cases better than for stars in our
Galaxy (KK1). This is probably due to the fact that the distance to SMC is
known with a relatively high precision and, thus, the basic stellar parameters
are known also with a high precision (probably with exception of {some
systematic effects like} wind
blanketing effect which may influence parameters of both stellar groups).
 However, there is a significant
disagreement between theoretical and observ{ed} values for lower mass-loss
rates ($\dmdt \lesssim10^{-7}\,\smrok)$. In this case the predicted
mass-loss rates are more than ten times higher than that derived from
spectral analysis of observed data. This is not only a problem of our
models, since also predictions of VKL show the
same behaviour (see B03; Martins et al. {\citeyear{martin,okali}}). To
demonstrate this
conclusion, we have {added} the mass-loss rate predictions of studied
stars calculated using VKL recipe {into} Fig.~\ref{dmdt}. {The possible origin of this discrepancy will
be discussed in Sect.~\ref{trenikap}.}

\zm{Our results for possible young Vz stars with thin winds ($\dmdt
\lesssim10^{-7}\,\smrok)$ from Mr04 sample do not show that the systematic 
disagreement between predicted mass-loss rates and mass-loss rates derived from
observation is higher than for generally older stars with thin winds from B03 sample (see
Fig.~\ref{dmdt}).}

\zm{For stars with very small mass-loss rates ($\dmdt \lesssim10^{-7}\,\smrok)$
only upper limits of their terminal velocities are
available in the literature. Predicted terminal velocities of these stars are
consistent with these upper limits. Note however that this is not
a check of the models because the standard wind theory predicts that the
terminal velocities and mass-loss rates are related -- if the mass-loss rate of
these stars is really much lower than the prediction derived in this paper, then their
terminal velocities may be much higher.}

The mass-loss rate{s} derived by VKL recipe are
slightly higher than that of us although KK1 found relatively good agreement
between these theoretical rates. This is caused by {a} different
boundary flux used.
The line-blanketed fluxes from OSTAR2002 grid used in this work have lower flux
in the UV domain (this domain is important for {the} line-driving) than H-He models
used by KK1, consequently derived mass-loss rates are slightly lower. However, the
effect of different boundary fluxes on terminal velocities is small.

\subsection{Wind momentum luminosity relationship}

\begin{figure}
\resizebox{\hsize}{!}{\includegraphics{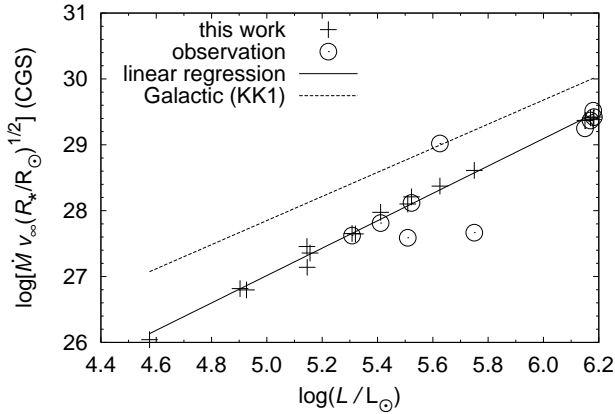}}
\caption[]{Comparison of calculated ({crosses}) and observed ({circles})
{modified} wind momentum for considered stars. Note that {observed values for}
stars with unknown terminal
velocity were excluded from this plot {since it is possible to obtain at most
 only
lower limits to their {modified} wind momentum} \zm{(for stars with known
observed mass-loss rates)}. Linear regression of calculated data 
of SMC and Galactic (KK1) stars is also plotted.}
\label{momentak}
\end{figure}

Wind momentum-luminosity relationship
(see Kudritzki \& Puls (\citeyear{kupul}) and references therein)
may provide an independent method for {the} determination of
stellar and consequently also galactic distances. However, to achieve this,
detailed calibration of this relationship is necessary, especially with respect
to {the} metallicity. {L}ow-metallicity environment of SMC provide{s} an ideal
tool for such a calibration.

We compare theoretical modified wind momentum-luminosity
relationship $ \dmdt v_\infty \zav{R_*/\text{R}_{\sun}}^{1/2}$
obtained for considered stars {using} our NLTE wind models with a relationship
derived from observations of these stars (see Fig.~\ref{momentak}). We excluded
{observed values for}
stars for which {only upper limits of} the{ir} terminal velocities {were
derived from observation}.
We conclude that there is a relatively good agreement between calculated
{values and those}
derived from observation. Note however that the agreement for excluded
stars with unknown terminal velocity {(for stars with
$\dmdt<10^{-7}\,\text{M}_\odot\,\text{year}^{-1}$)}
{is very poor (using either lower limits of their terminal velocities or
terminal velocities derived from our calculations)}, since their
predicted mass-loss rates are much higher than {the} observed ones.
\zm{On the other hand standard theory predicts that terminal velocities and
mass-loss rates are related. In such a case the terminal velocities of these
stars could be much higher and the agreement between predicted and observed wind
momentum-luminosity relationship could be improved.}

\begin{table}
\centering
\caption{Comparison of {modified} wind momentum-luminosity relationship (see Eq. (\ref{rovmomlumvztah})) {taken} from different sources. SMC denotes sample considered in this paper,
either calculated from theoretical values or from values derived from
observation {(again excluding the observed values for stars with very weak
winds)}.}
\label{wmlrtab}
\begin{tabular}{lcc}
\hline
\hline
Sample & $\log D_0$ & $x$ \\
\hline
Galactic (Vink et al. \citeyear{vikola}) & $18.68 \pm 0.26$ &$ 1.826 \pm 0.044$\\
Galactic (KK1) & $18.7\pm2.3$ & $1.83\pm0.40$\\
SMC (theoretical) & ${16.6\pm0.2}$ & ${2.08\pm0.04}$\\
SMC (observed) & $16.6\pm2$ & $2.1\pm0.4$\\
\hline
\end{tabular}
\end{table}

We have also calculated the linear regression of both theoretical and
observ{ed} {modified} wind momentum-luminosity relationship for considered stars
\begin{equation}
\label{rovmomlumvztah}
\log\hzav{\dmdt v_\infty\zav{R_*/\text{R}_{\sun}}^{1/2}}=x\log(L/\text{L}_\odot)+\log D_0,
\quad \text{(CGS)}
\end{equation}
and compared it with the theoretical values derived for Galactic stars
{(Tab.~\ref{wmlrtab})}.
First, due to relatively good agreement of mass-loss rates there is a good
agreement between theoretical and observ{ed} {modified} wind momentum-luminosity
relationship for SMC stars {(exluding stars with thin wind)}. Moreover, due to lower mass-loss rate{s} of SMC stars the $D_0$ value
is significantly lower than {that} for Galactic stars. Also the slope is slightly
different. {Roughly $5-6\times$ lower modified wind momentum derived in the
present study compared to that derived by KK1 for Galactic stars is mostly
due to lower wind mass-loss rates caused by low metallicity of SMC and to a
lesser extent by a
downward revision of mass-loss rates due to the use of blanketed model
atmospheres.}

\section{Influence of the uncertainties of stellar parameters determination}

{%
Individual stellar parameters (like {the} effective temperature, mass or
radius) \zm{are not in some cases}
determined with a high degree of precision. This is especially true for
hot stars for which large uncertainties of derived \zm{stellar radii (for
Galactic stars due to uncertainties of the determination of their distance)}
and stellar masses
(see Herrero et al. \citeyear{hekuku}, Lanz et al. \citeyear{hmotak}) \zm{may} 
exist. 
}

{The uncertainties of determination of the stellar parameters influence also
the predicted wind parameters (i.e.~the
mass-loss rates and the terminal velocities). To describe this in detail,
we perform study of variations of
predicted wind parameters with small change of stellar parameters. From the
scaling of modified CAK theory (Kudritzki et al. \citeyear{kustar})
\begin{equation}
\label{kudmdt}
\dot M\sim N_0^{1/\alpha'}L^{1/\alpha'}\hzav{M\zav{1-\Gamma}}^{1-1/\alpha'},
\end{equation}
where $N_0$ is connected with number of lines that effectively drive the
stellar wind, $\Gamma$ is the Eddington parameter and
\begin{equation}
\alpha'=\alpha-\delta,
\end{equation}
where $\alpha$ and $\delta$ are usual CAK parameters,  so $\alpha'$ is
typically equal to about $0.5$. Hot star wind mass-loss
rate mostly depends on the stellar luminosity $L$. With decreasing
metallicity parameter $N_0$ decreases causing a decrease of mass loss
rate. Finally, with decreasing effective mass $M\zav{1-\Gamma}$  mass loss rate
increases. Since Eq.~(\ref{kudmdt}) does not take into account the variations of
$\alpha'$ with e.g.~the effective temperature, we also include the scaling of Vink
et al. (\citeyear{vikola}) obtained for Galactic OB stars
\begin{equation}
\label{vkldmdt}
\dot M\sim L^{2.2}M^{-1.3}\Teff\zav{v_\infty/v_\text{esc}}^{-1.3},
\end{equation}
with result of metallicity dependence derived by VKL as
\begin{equation}
\label{vklz}
\dmdt\sim Z^{0.69}.
\end{equation}
From the classical CAK theory it is known that the terminal velocity $v_\infty$
depends mostly on the escape velocity $v_\text{esc}$ (see Kudritzki
\& Puls (\citeyear{kupul}) for a review).
}

{Here we do not aim to fit expressions like Eq.~(\ref{kudmdt}) for large
amount of stars, however just to study variations of wind parameters with
varying stellar parameters.}

\subsection{Effective temperature and stellar mass}

\begin{figure*}
\resizebox{0.40\hsize}{!}{\includegraphics{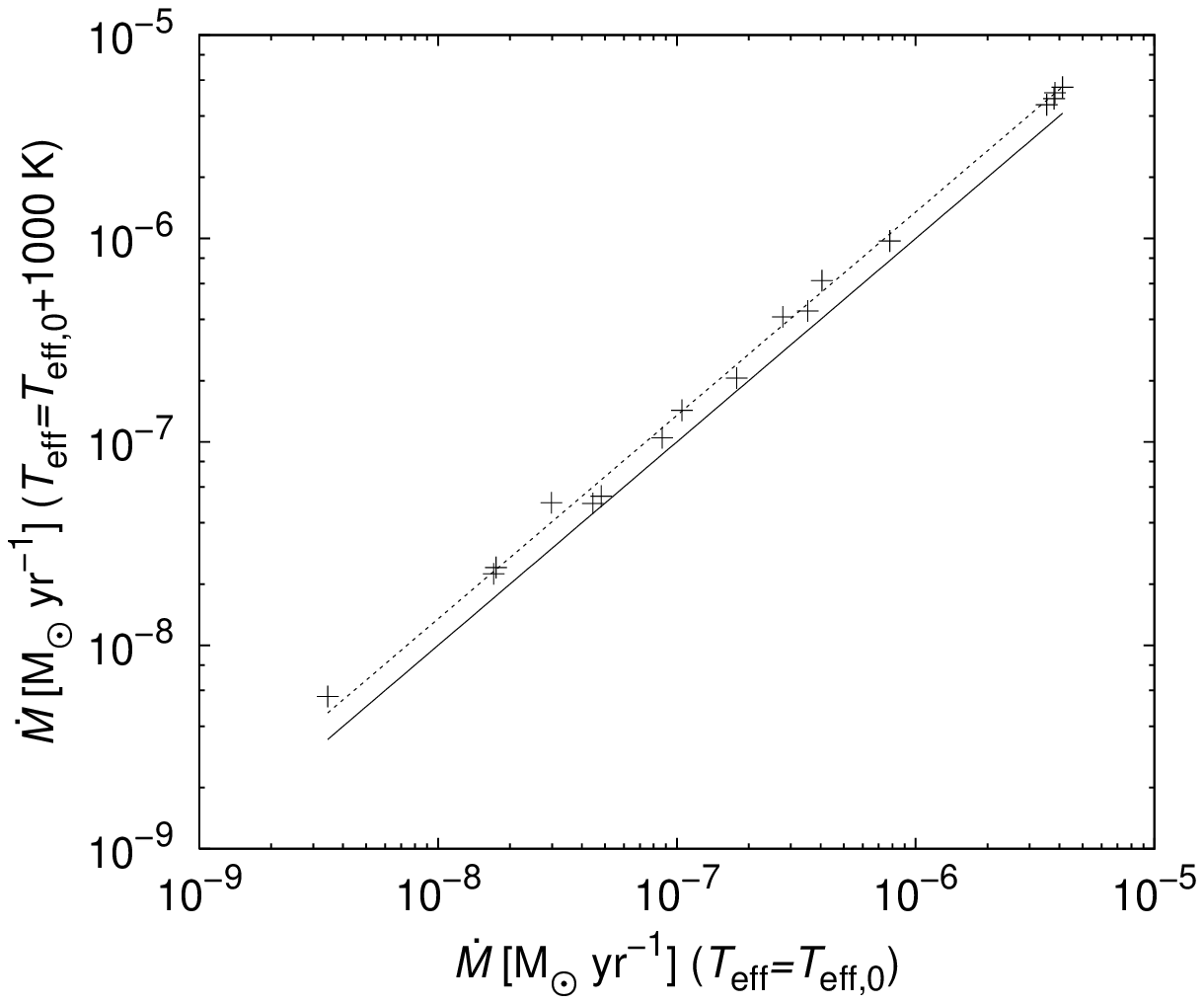}}
\resizebox{0.40\hsize}{!}{\includegraphics{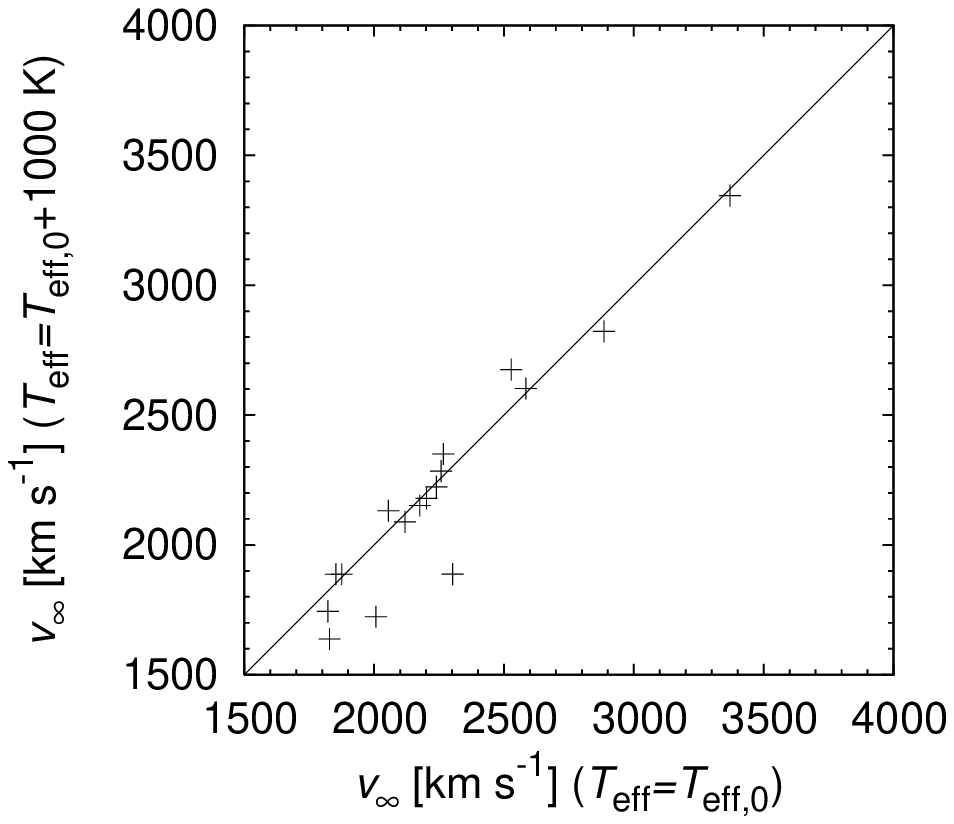}}
\caption[]{{Comparison of {the} mass-loss rates ({\em left panel}) and
{the} terminal
velocities ({\em right panel}) of studied stars calculated with
original effective temperature and effective temperature higher by
$1000\,\text{K}$. Solid line denotes one
to one relation and dashed line denotes linear fit to the relation between two
groups of wind parameters.}}
\label{tdmdtvnek}
\end{figure*}

\begin{figure*}
\resizebox{0.40\hsize}{!}{\includegraphics{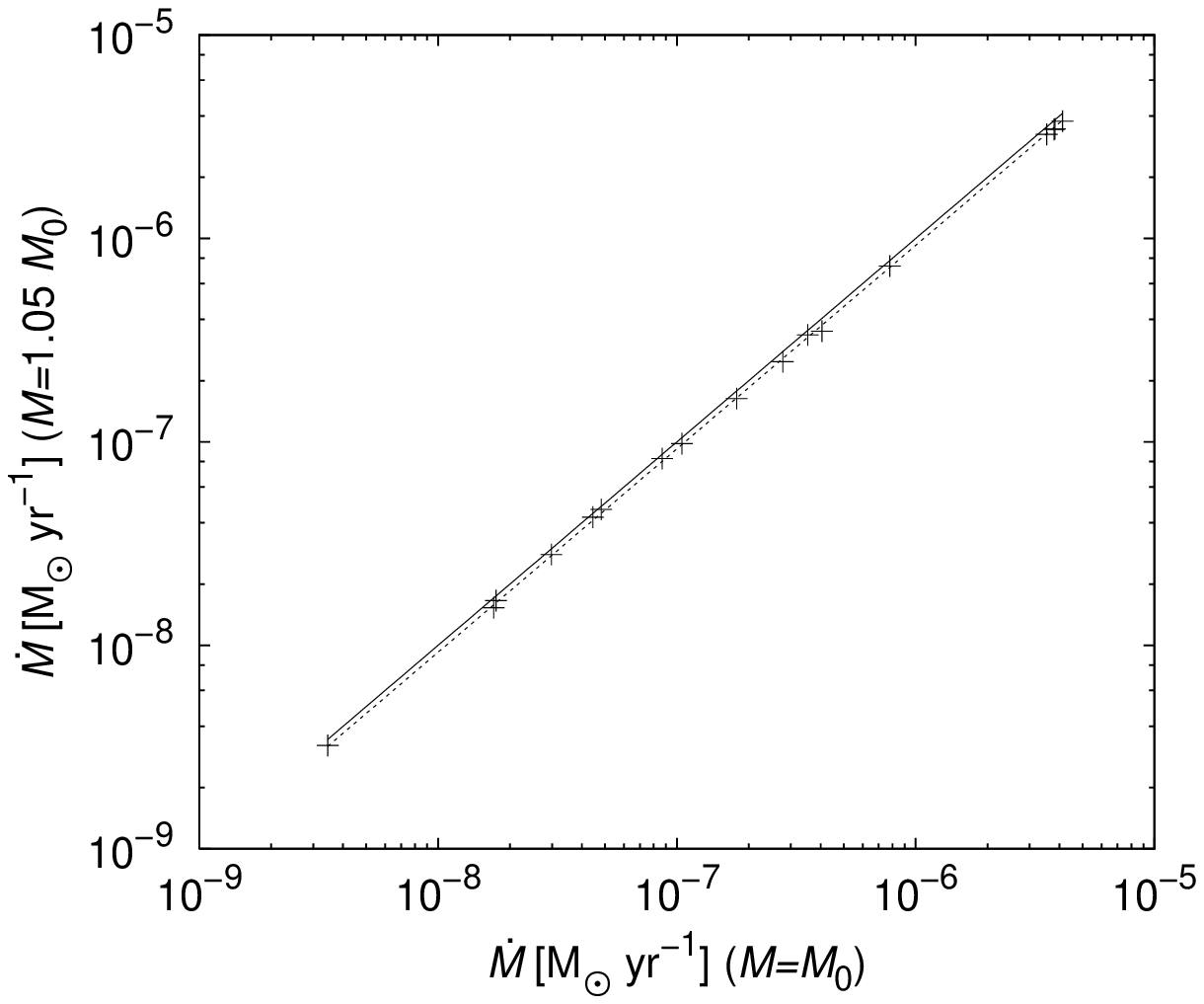}}
\resizebox{0.40\hsize}{!}{\includegraphics{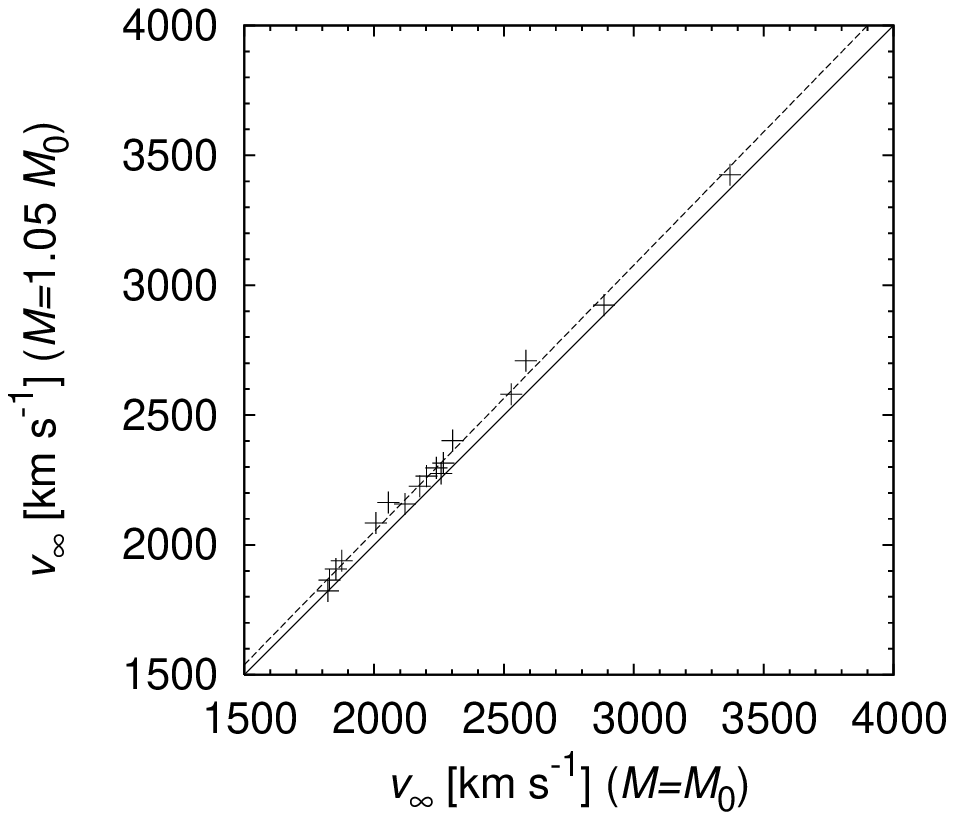}}
\caption[]{{Comparison of {the} mass-loss rates ({\em left panel}) and
{the} terminal
velocities ({\em right panel}) of studied stars calculated with
original mass and mass $1.05$ times higher. Solid line denotes one
to one relation and dashed line denotes linear fit to the relation between two
groups of wind parameters.}}
\label{mdmdtvnek}
\end{figure*}


Comparison of mass-loss rates and terminal velocities for studied stars
calculated with original set of parameters and with stellar effective
temperature by $1000\,\text{K}$ higher is given in Fig.~\ref{tdmdtvnek}.
Apparently, 
for higher effective temperature the mass-loss rate is higher, in agreement with
other theoretical predictions (see Eqs.~(\ref{kudmdt}), (\ref{vkldmdt})).
From our calculations it is possible to
derive that the mass-loss rate scales with the effective temperature as
\begin{equation}
\dmdt\sim T_\text{eff}^{10.64},
\end{equation}
from which the effective parameter $\alpha^\prime=0.38$. 
{This scaling is in a
relatively good agreement with scaling of Vink et al. (\citeyear{vikola}, see
Eq.~(\ref{vkldmdt})), which implies $\dmdt\sim T_\text{eff}^{9.8}$.} The
dependence of {the} terminal velocity on the effective temperature
(Fig.~\ref{tdmdtvnek}) is more scattered.

The variations of the mass-loss rate and the terminal velocity with stellar mass
are more stringent (Fig.~\ref{mdmdtvnek}). With increasing stellar mass the mass loss rate decreases,
on average 
\begin{equation}
\dmdt\sim M^{-1.60},
\end{equation}
{with fair agreement with  Eqs.~(\ref{kudmdt}), (\ref{vkldmdt}) for the
derived value $\alpha^\prime=0.38$}.
The terminal velocity is proportional to the escape velocity.
From our calculations we derive the average relation
\begin{equation}
v_\infty\sim M^{0.52},
\end{equation}
which clearly reflects {this proportionality.}

It is clear that some part of the discrepancy between observed and theoretical
wind parameters may be attributed
to the uncertainties in the determination of the stellar parameters (mass and
{the} effective temperature). {Especially the downward revision of the
stellar mass by the factor of $1.5\times$ (e.g.~due to the difference between
spectroscopic and evolutionary masses) would cause downward revision of the
terminal
velocities by the factor of about $1.3$ and the increase of the mass-loss rate
by $2\times$.} On the
other hand, since the distance to the SMC is known with a relatively high degree
of precision we conclude that probably also another source of discrepancy between
observed and predicted wind terminal velocities is present. Moreover, if higher
terminal velocities scatter is given purely by the uncertainties of determination
of stellar mass and effective temperature, then similar
effect should be also present in KK1, where slightly better agreement between
theoretical and observed terminal velocities was derived.

\begin{figure*}
\resizebox{0.40\hsize}{!}{\includegraphics{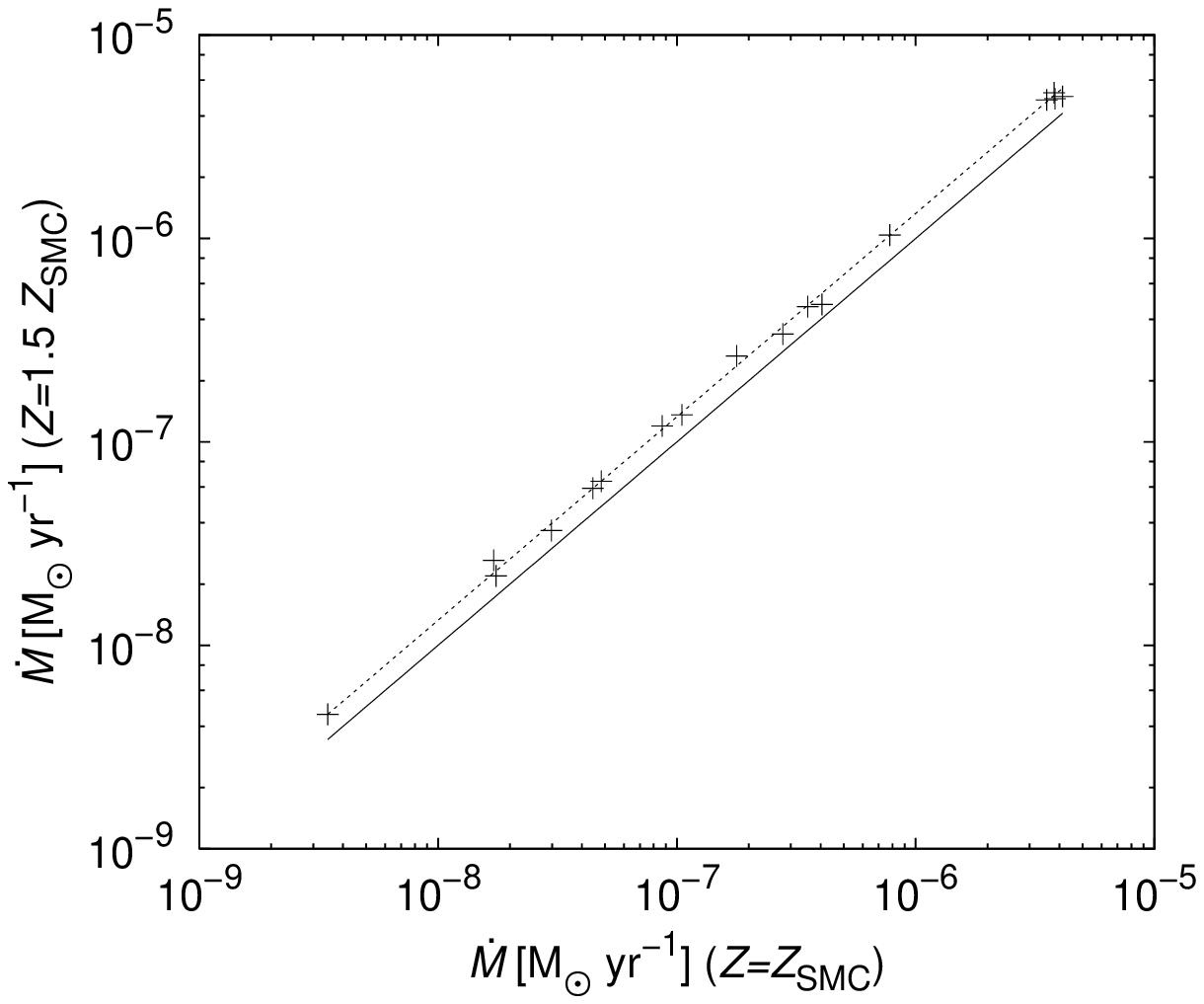}}
\resizebox{0.40\hsize}{!}{\includegraphics{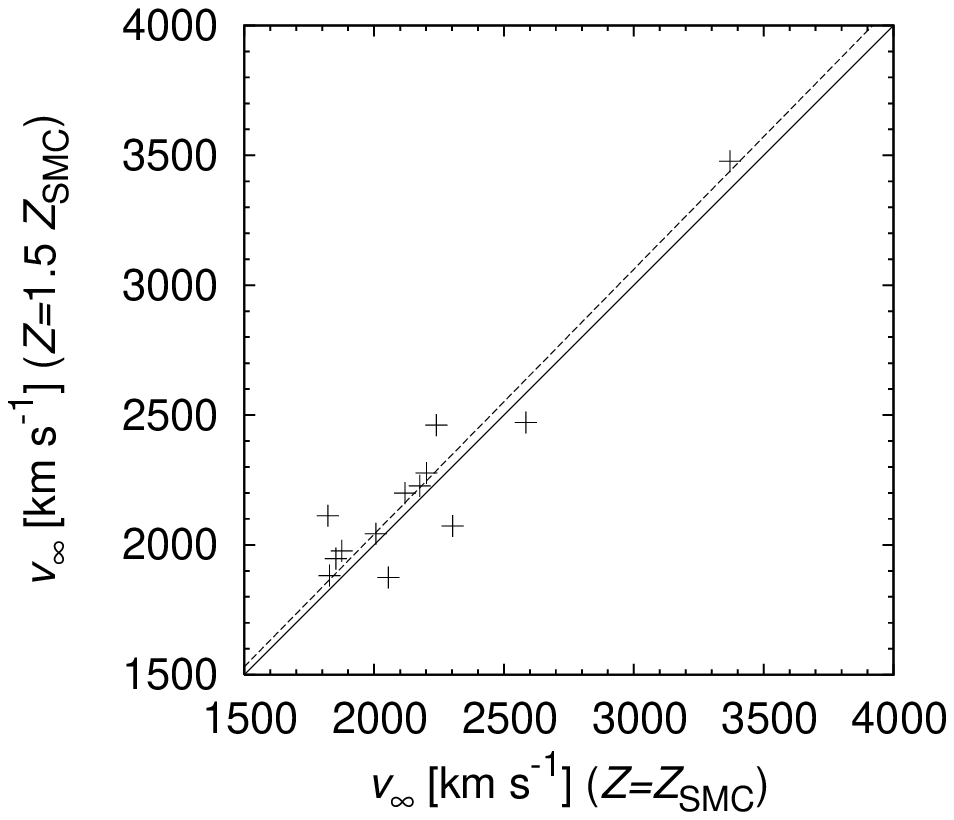}}
\caption[]{Comparison of {the} mass-loss rates {(\em{left panel})} and
{the} terminal
velocities {(\em{right panel})} of studied stars calculated with
original metallicity and metallicity $1.5$ times higher. Solid line denotes one
to one relation and dashed line denotes linear fit to the relation between two
groups of {wind parameters}.}
\label{zdmdtvnek}
\end{figure*}

\subsection{Abundances}
\label{kapvlivz}



To study the variations of wind parameters with metallicity we recalculated wind
models with the same stellar parameters as in Tab.~\ref{mmmhvezpar}, however
with abundance of {heavier elements} $1.5$ times higher. Comparison of {the} mass-loss
rates and {the} terminal velocities calculated with different metallicities is given
in Fig.~\ref{zdmdtvnek}.

{F}or higher metallicity the {radiative force is higher and
consequently also the} mass loss rate is higher. We have found that
for studied stars the relation
\begin{equation}
\dmdt\sim Z^{0.67}
\end{equation}
holds. This is in a good agreement with {VKL} {(see
Eq.~(\ref{vklz}))}.

The situation with the variations of {the} terminal velocity with the metallicity is
more complicated. In Sect.~\ref{srovnavacka} we concluded that the ratio of the
wind terminal velocity to {the} stellar escape velocity $v_\infty/v_\text{esc}$
{may be slightly} higher for Galactic stars than for SMC stars. 
With varying metallicity the
wind density changes, consequently wind ionization {and excitation}
state varies,
{and hence} also terminal velocit{ies} of studied stars var{y}. However, these
variations are not monotonic, as can be seen from Fig.~\ref{zdmdtvnek}. For some
stars the terminal velocity does not significantly change with metallicity, for
some of them increases, for some of them decreases. On average, the {terminal
velocity} slightly increases with increasing metallicity for
studied SMC stars{,}
\begin{equation}
v_\infty\sim Z^{0.06}.
\end{equation}
{From calculations of Kudritzki (\citeyear{kudmet}) we can infer more steep
proportionality, roughly $v_\infty\sim Z^{0.12}$}.

\begin{figure}
\resizebox{\hsize}{!}{\includegraphics{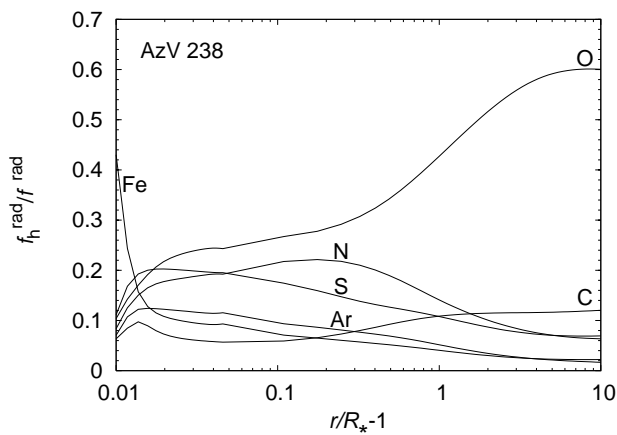}}
\caption[]{{The radial variation of {the} contribution of the radiative force $f_{h}^\text{rad}$
acting on element~$h$ to the total radiative force $f^\text{rad}$  for star
AzV~238 (plotted as $w_{h}^\text{rad}=f_{h}^\text{rad}/f^\text{rad}$).
Iron is the 
most important element for the radiative {driving} close to the star, while oxygen
dominates in the outer regions.}}
\label{av238grel}
\end{figure}

\begin{figure*}
\resizebox{0.49\hsize}{!}{\includegraphics{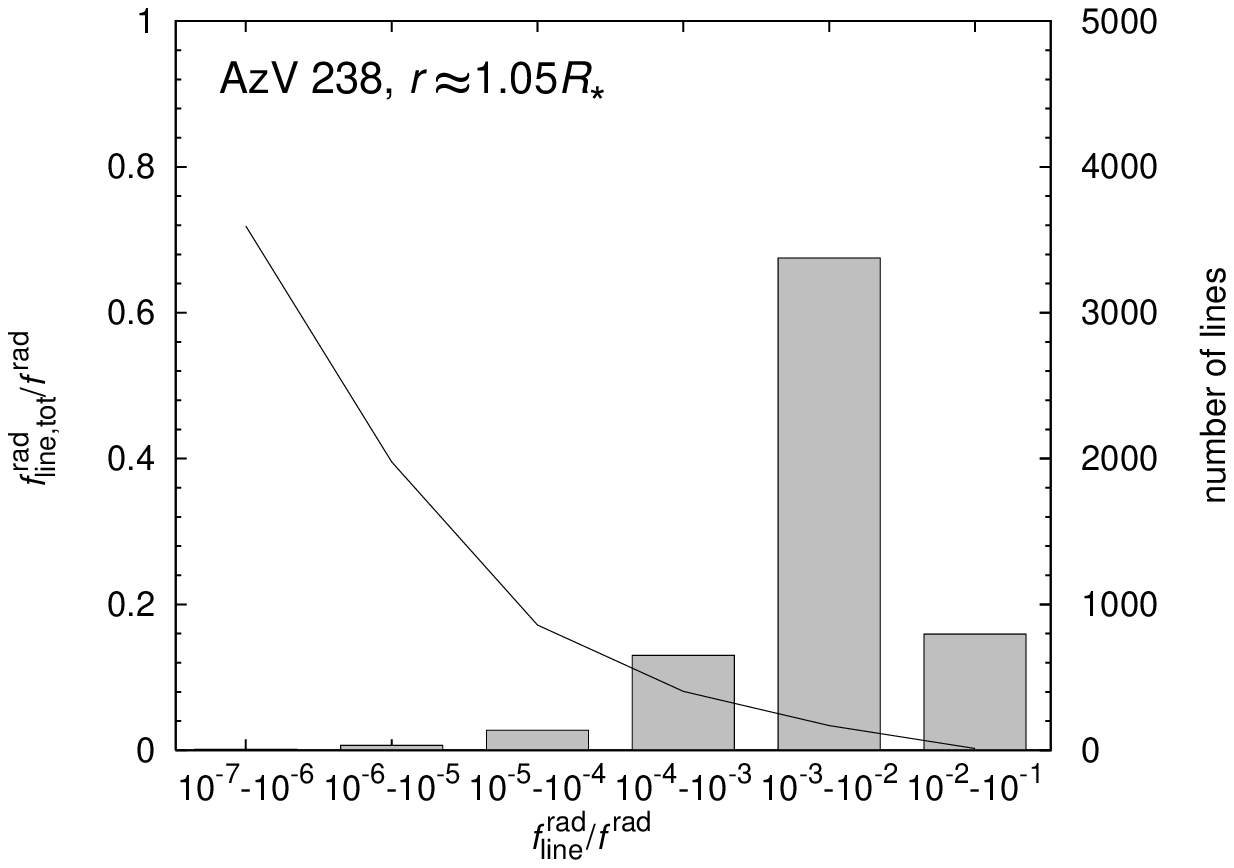}}
\resizebox{0.49\hsize}{!}{\includegraphics{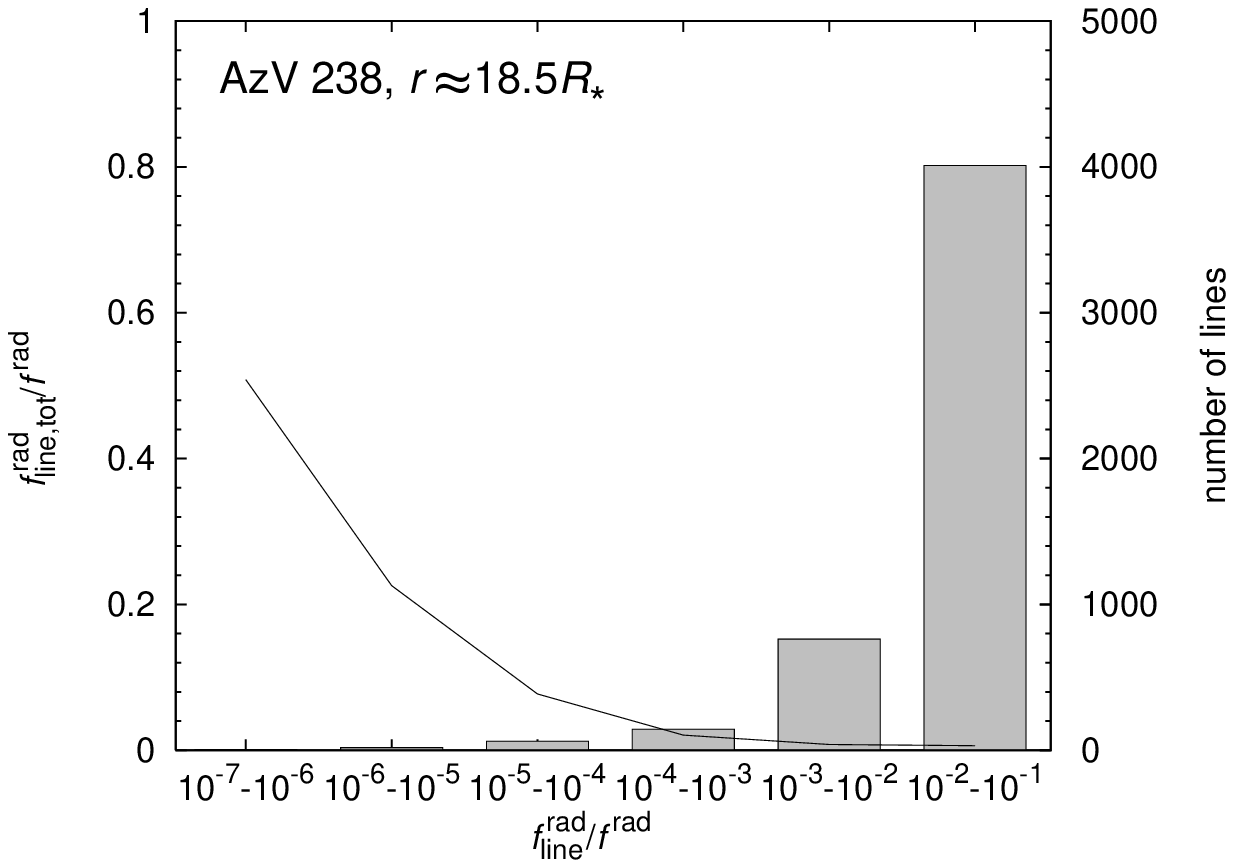}}
\caption[]{{Contribution of lines with different strengths to the radiative
force at the critical point ($r\approx1.05R_*$, {\em left panel}) and in the
outer wind region ($r\approx18.5R_*$, {\em right panel}) for star AzV~238. The total relative
contribution to the radiative force $f_\text{line, tot}^\text{rad}/f^\text{rad}$
\zm{summed over} lines with given \zm{individual} contribution to the radiative force
$f_\text{line}^\text{rad}/f^\text{rad}$ is plotted using boxes. Solid line
denotes number of spectral lines whose relative contribution to the radiative
force lies in a given interval. Radiative acceleration close to the star is
mainly due to hundreds of lines whose $f_\text{line}^\text{rad}/f^\text{rad}$ lies
in the interval $(10^{-3}-10^{-2})$, whereas the acceleration in the outer
regions is given just by few dozens lines with
$10^{-2}<f_\text{line}^\text{rad}/f^\text{rad}<10^{-1}$.
}}
\label{av238car}
\end{figure*}

\subsection{Sensitivity of $v_\infty$ on the conditions in the outer wind}
\label{citlivka}

{Although {\em on average} the value of $v_\infty$ depends on metallicity
only slightly, the high scatter between terminal velocities derived using
different metallicities in Fig.~\ref{zdmdtvnek} may seem surprising. Similar effect was however, in
a different context, reported also by other authors (e.g.~Puls et al.
\citeyear{pusle}). The O star wind{s} in the outer regions {are} accelerated 
mainly by few {dozen} lines of lighter elements (like C, N, O, see Pauldrach \citeyear{pasam}, Vink et al.
\citeyear{vikolabis} and also Figs.~\ref{av238grel} and~\ref{av238car}). In such a
case, the line acceleration is very sensitive to {the}
detailed wind structure (i.e. temperature, density and chemical composition).
{Partly} due to this, large variations of $v_\infty/v_\text{esc}$ (both observed and
theoretical, see Lamers et al. \citeyear{lsl} and Pauldrach at al. \citeyear{pavyvoj})
occur for individual stars. This is also likely cause for the high scatter
between terminal velocities derived using       different metallicities in
Fig.~\ref{zdmdtvnek}. This scatter also occurs when variations of terminal
velocity with temperature (see Fig.~\ref{tdmdtvnek}) are studied. To a lesser
extent this is also true for the variations of stellar mass, see
Fig.~\ref{mdmdtvnek}.}

The metallicity of studied stars is not known with {a} high precision. For
\zm{many} stars we used just scaled Galactic chemical composition, although
significant deviations from this scaling exist ({e.g.}~B03; they are also probably
manifested {by} a different number ratio of WN to WC stars in the Clouds, see
Mikul\'a\v sek \citeyear{mik}). Moreover, we
do not have any information on the abundance of many elements which are 
important for {the} radiative driving at all (e.g. Ne, Ar). Even worse, there are
large differences in the chemical composition of individual stars ({e.g.}~B03). To
conclude, {some} part of higher scatter between observed and predicted terminal
velocities can be attributed to poorly known chemical composition of most of the
studied stars. {Uncertainties of the determination of other stellar
parameters, i.e. the stellar mass and {the} effective temperature may
\zm{(besides the approximations involved in our code)} also contribute
to this scatter.}

\section{Multicomponent effects}
\label{trenikap}

Stellar wind{s} of hot stars {are} accelerated mainly by the absorption of radiation
in the resonance lines of heavier elements. The radiative acceleration acting on
individual elements is however different, and consequently individual elements
have different velocities. {Consequently}, stellar wind{s} of hot stars have a
multicomponent nature ({e.g.~Springmann \& Pauldrach \citeyear{treni},} KKII). For many stars the velocity differences are
small, thus the wind multicomponent nature can be neglected in this case. 
Krti\v cka et al. (\citeyear{gla}) showed that in the low-metallicity environment
the velocity differences between wind components are larger, thus
multicomponent effects are more important. However, as was noted e.g. by
{Mr04}, even
for relatively low-metallicity environment of stellar wind of SMC stars the multicomponent effects can be
neglected.

On the other hand, these considerations are based on the assumption that the heavier
elements can be described by one component only. By other words, usually it is
assumed that {atomic} mass of heavier ions is the same, the radiative
acceleration on heavier elements is the same and consequently these elements
have the same {mean} velocity. However, in reality individual elements have different
mass, the radiative acceleration acting on them is {also} different and consequently
these elements have different {mean} velocities. {Hence}, multicomponent effects may
occur for higher densities and metallicities than previously assumed.

\subsection{Velocity differences in the stellar wind}

To estimate the importance of multicomponent effects in the stellar wind of SMC
stars we calculate approximate velocity differences between individual
{heavier} elements
$h$ {(accelerated by the line absorption)}
and passive wind component p (hydrogen and helium) using the momentum equation of
individual wind elements. In the case of stationary spherically symmetric stellar
wind this equation reads (cf. Burgers \citeyear{burgers}, KKII)
\begin{multline}
\label{rovhyb}
{\vr}_a\frac{\text{d} {\vr}_a}{\text{d}r}=
{g}_{a}^{\mathrm{rad}}-g-\frac{1}{{\rho}_a}\frac{\text{d}}{\text{d}r}\zav{{a}_a^2{\rho}_a}
 +\frac{q_a}{m_a}E+ \\*
                +\frac{1}{{\rho}_a}
\sum_{b\neq a} K_{ab}G(x_{ab})\frac{{\vr}_b-{\vr}_a}{|{\vr}_b-{\vr}_a|},
\end{multline}
where ${\vr}_a$ and ${\rho}_a$ are the velocity and the density of component $a$
 {($a=\text{p}$ or $a=h$)},
${a}_a$ is {the} isothermal sound speed, $E$ is {the} charge separation electric
field, $g$ is {the} gravitational acceleration, ${g}_{a}^{\mathrm{rad}}$ is the
radiative acceleration {and $q_a$ and $m_a$ are charge and mass of particle
$a$}. {Frictional parameter} has the following form:
\begin{equation}
{K}_{ab}={n}_a{n}_b\frac{4\pi  {q}_a^2{q}_b^2}{k T_{ab}}\ln\Lambda,
\end{equation}
where ${n}_a$ and ${n}_b$ are number densities of individual components
and mean temperature of both components {is}
\begin{equation}
T_{ab}=\frac{m_aT_b+m_bT_a}{m_a+m_b}
\end{equation}
{($T_a$ and $T_b$ are temperatures of wind components $a$ and $b$)}
and $\ln\Lambda$ is the Coulomb logarithm. The argument of the Chandrasekhar
function $G(x_{ab})$ is
\begin{equation}
x_{ab}=\frac{|{\vr}_b-{\vr}_a|}{\alpha_{ab}},
\end{equation}
where
\begin{equation}
\alpha_{ab}^2=\frac{2k\zav{m_aT_b+m_bT_a}} {m_am_b}.
\end{equation}
For low velocity differences ($x_{ab}\lesssim1$) the flow is well coupled.
However, for higher velocity differences ($x_{ab}\gtrsim1$) the Chandrasekhar
function is decreasing and this behaviour {may enable} dynamical decoupling of wind
components {(see Springmann \& Pauldrach \citeyear{treni}, KKII)}.

In the momentum equation of heavier ions {Eq.~(\ref{rovhyb})} the left-hand side term and the
pressure {term} can be neglected. Also the gravitational acceleration and {the} electric
polarisation field term can be neglected. Finally, due to high number density
of passive component (hydrogen and helium) compared to {the} number
density of heavier elements the only important frictional term is that between
passive and other components. Consequently, the approximative
momentum equation of heavier element $h$ is
\begin{equation}
\label{rovhybprib}
{g}_{h}^{\mathrm{rad}}= \frac{1}{{\rho}_h}
K_{h\text{p}}G(x_{h\text{p}}).
\end{equation}
{T}his equation states that whole momentum {obtained} by \zm{individual} heavier
elements due to the line-absorption is transferred by friction to {the} passive
component. Using the {Taylor expansion of the} Chandrasekhar function 
({appropriate}
for $x_{h\text{p}}<1$) $G(x_{h\mathrm{p}})\approx\frac{2
x_{h\mathrm{p}}}{3\sqrt{\pi}}$ the relative velocity difference between passive
component and a given heavier ion $h$ is
\begin{equation}
x_{h\text{p}}=\frac{|{\vr}_\text{p}-{\vr}_{h}|}{\alpha_{h\text{p}}}\approx
{g}_{h}^{\mathrm{rad}}\frac{m_{h}}{{n}_\text{p}}
\frac{3kT}{8\sqrt\pi  {q}_{h}^2{q}_\text{p}^2\ln\Lambda},
\end{equation}
where we have assumed $T_\text{p}\approx T_h\approx T$.
For our discussion the relevant quantity is not the radiative acceleration
${g}_{h}^{\mathrm{rad}}$
itself, however the radiative force
${f}_{{h}}^{\mathrm{rad}}=\rho_{h}{g}_{h}^{\mathrm{rad}}$ (per
unit volume).
Thus,
\begin{equation}
\label{berlin}
x_{h\text{p}}\approx
{f}_{{h}}^{\mathrm{rad}}\frac{1}{{n}_{h}{n}_\text{p}}
\frac{3kT}{8\sqrt\pi  {q}_{h}^2{q}_\text{p}^2\ln\Lambda}.
\end{equation}
The work done by the radiative acceleration is used to lift the wind material
from the stellar gravitational well. Thus, for stars with the same mass-loss
rates {(and the same velocity field)} the radiative force (at corresponding radii) shall be similar.  
{These stars with} lower metallicities have lower
${n}_{h}$ and {consequently} higher $x_{h\text{p}}$. In reality, stars with lower
metallicities have also lower mass-loss rates, lower ${n}_\text{p}$
and {consequently} even higher
$x_{h\text{p}}$. Due to these two effects  (see also Krti\v cka et al.
\citeyear{gla}) stars with lower metallicities have higher velocity differences
\zm{between wind components}.
The crucial point is, however, that heavier {elements} which are not abundant in the
stellar wind (i.e. ${n}_{h}\ll{n}_\text{p}$) and which significantly
contribute to the radiative acceleration (i.e. their
${f}_{h}^{\mathrm{rad}}$ is large) may have large relative velocity
differences $x_{h\text{p}}$, in many cases $x_{h\text{p}}\approx1$. In such a
case, \zm{instability connected with the} decoupling of considered element may occur.

Let us first roughly estimate in which situation this may occur. Let be
$w_{h}^{\mathrm{rad}}$ {the} relative contribution of a given element $h$
to the total radiative force ${f}^{\mathrm{rad}}$, i.e. 
\begin{equation}
\label{wittenberge}
{f}_{h}^{\mathrm{rad}}=w_{{h}}^{\mathrm{rad}}{f}^{\mathrm{rad}}.
\end{equation}
{Neglecting the gravity, the pressure term and the electric polarisation
field} the total radiative force
{can be approximated from the momentum equation Eq.~(\ref{rovhyb}) of passive
component as}
\begin{equation}
{f}^{\mathrm{rad}}\approx\pr\rho{\vr}\frac{\text{d} {\vr}}{\text{d}r},
\end{equation}
{where $\vr$ is the mean wind velocity, we may assume $\vr\approx\pr\vr$.}
The velocity gradient can be estimated as 
\begin{equation}
\label{gradry}
\vr\frac{\text{d} {\vr}}{\text{d}r}\approx\frac{v_\infty^2}{R_*},
\end{equation}
for $\ha n$ we can write {from the approximate continuity equation}
\begin{equation}
\label{ludwiglust}
\ha n\approx\ha Z\frac{\dmdt}{4\pi R_*^2v_\infty\ha m},
\end{equation}
where $\ha Z$ is the {density} of a given element relative to the bulk density
$\rho$ in
the stellar atmosphere ($\ha\rho=\ha Z\rho$). 
Consequently using Eqs.~\eqref{berlin}--\eqref{ludwiglust} we derive
\begin{equation}
\label{ludwiglust2}
x_{h\text{p}}\approx w_{{h}}^{\mathrm{rad}}{v_\infty^3}{R_*} 
\frac{\pr m\ha m}{\ha Z\dmdt}
\frac{3\sqrt\pi kT}{2  {q}_{h}^2{q}_\text{p}^2\ln\Lambda}.
\end{equation}
In the case when the {heavier} ions are approximated by one component only, we
have $w_{{h}}^{\mathrm{rad}}=1$ and we arrive at Eq.~(6) of Krti\v cka
et al. (\citeyear{gla}). However, in many cases {the abundance of individual
element} $\ha Z$ is much lower than $\xa$
{(the ratio of total density of heavier "absorbing" ions and passive component in the
atmosphere)} and
thus, predicted velocity differences are larger. Eq.~(\ref{ludwiglust2}) can be
also rewritten in a more convenient form as
\begin{equation}
\label{xskal}
x_{h\text{p}}\approx 0.015 w_{{h}}^{\mathrm{rad}} v_8^3 R_{12}
\frac{\ha A}{\ha Z\dmdt_{-11}}
\frac{T_4}{{z}_{h}^2{z}_\text{p}^2},
\end{equation}
where we have assumed $\pr m=m_\text{H}$, ${q}_\text{p}=e{z}_\text{p}$,
${q}_{h}=e{z}_{h}$ and $\ha m=\ha A m_\text{H}$,
where $e$ is the elementary charge and $m_\text{H}$ {is the} proton mass,
and scaled quantities are ${\dot M}_{-11} \equiv {\dmdt}/(10^{-11}
\text{M}_\odot\,\mathrm{yr}^{-1})$, $v_{8}\equiv v_\infty/(10^8
\mathrm{cm}\,\mathrm{s}^{-1})$,
$R_{12}\equiv R_*/(10^{12}\mathrm{cm})$, and $T_4\equiv T/(10^4\text{K})$. 
{From Eq.~(\ref{xskal}) it follows that} for heavier
elements ($\ha A\approx10$) with low abundance ($\ha Z\approx10^{-5}$) 
which significantly contribute to the radiative force (i.e.~$\ha
w\approx 0.1$) the decoupling ($x_{h\text{p}}\gtrsim1$) can occur for relatively
large mass-loss rates (of order $10^{-8}\text{M}_\odot\,\mathrm{yr}^{-1}$).
{T}he
mass-loss rate for which the decoupling occurs in the winds of SMC stars is
approximately five times higher than the mass-loss rate for which the decoupling
occurs in the winds of Galactic stars, since
$Z_{h,\text{SMC}}\approx0.2Z_{h,\odot}$.

{Eq.~(\ref{xskal}) provides only a very approximate expression for
the velocity difference, mainly due to a very simplified velocity gradient
assumed in Eq.~(\ref{gradry}). For a more reliable estimate it is much better to
use Eq.~(\ref{berlin}).}

{Note that this picture of decoupling is somehow different from that
presented by other studies, i.e. Springmann \& Pauldrach (\citeyear{treni}) and KKII
and discussed by {Mr04}. These studies assumed that all
heavier
ions which are accelerated by line-transitions decouple from the mean (passive)
flow. However now we discuss a more realistic description that individual ions decouple from
the mean flow separately.}

\subsection{Velocity differences in the winds of studied SMC stars}

We used our NLTE wind code to calculate approximative velocity differences in
studied SMC {stellar winds}. For this purpose we {applied} Eq.~\eqref{berlin}, where both
the contribution of a given element to the radiative force $w_{{h}}$ and
charges of wind components $\ha q$ and $\pr q$ are calculated using our NLTE
wind code.

Calculated values of $x_{h\text{p}}$ for individual stars and individual
elements are given in Fig.~\ref{xpiobr}. Generally, the relative velocity
differences are smallest close to the star, where the stellar wind is
relatively dense. As the stellar wind accelerates, the wind density is lower and
velocity differences are higher. At some point the velocity differences have
its maximum and {for larger radii} decreas{e} outwards {due to decreasing
velocity gradient}. This general behaviour of relative
velocity differences was described elsewhere (KKII). In our case the behaviour is more complicated
mainly due to the processes of ionization and recombination.

{F}or most of the stars and
for most of the elements the velocity differences are rather low,
$x_{h\text{p}}\ll1$, thus there occurs no decoupling in this case. However, 
{in some cases, especially in those where the wind density is low,}
the relative velocity difference between argon (or sulphur)
and passive component
becomes higher and close to one. This is caused by the fact that argon and sulphur
have very low
metallicity ($Z_\text{Ar, SMC}\approx3\,10^{-5}$) and significantly contribute to
the radiative force ($w_\text{Ar}^\text{rad}\approx0.1$). 
The velocity differences are so high, that they may cause large frictional
heating. 

\subsection{Multicomponent wind models}

\begin{figure*}
\centering
\resizebox{0.31\hsize}{!}{\includegraphics{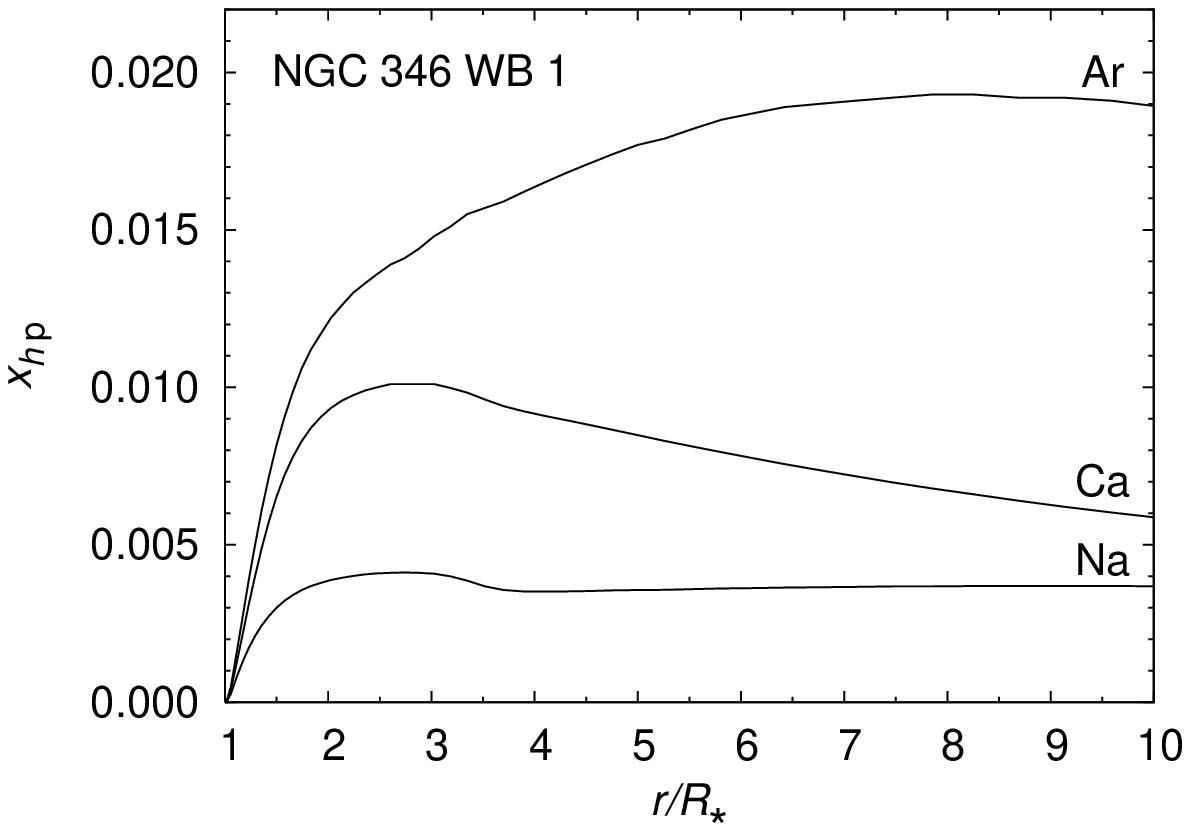}}
\resizebox{0.31\hsize}{!}{\includegraphics{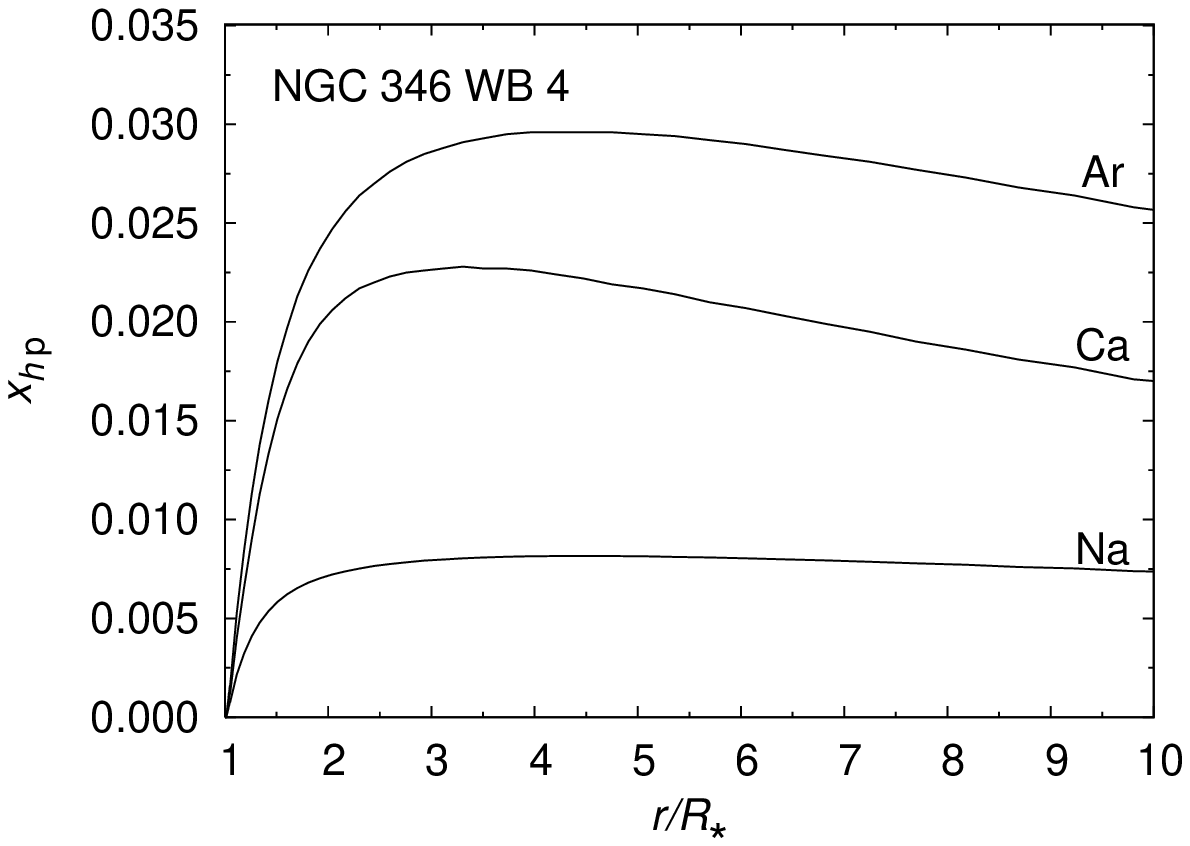}}
\resizebox{0.31\hsize}{!}{\includegraphics{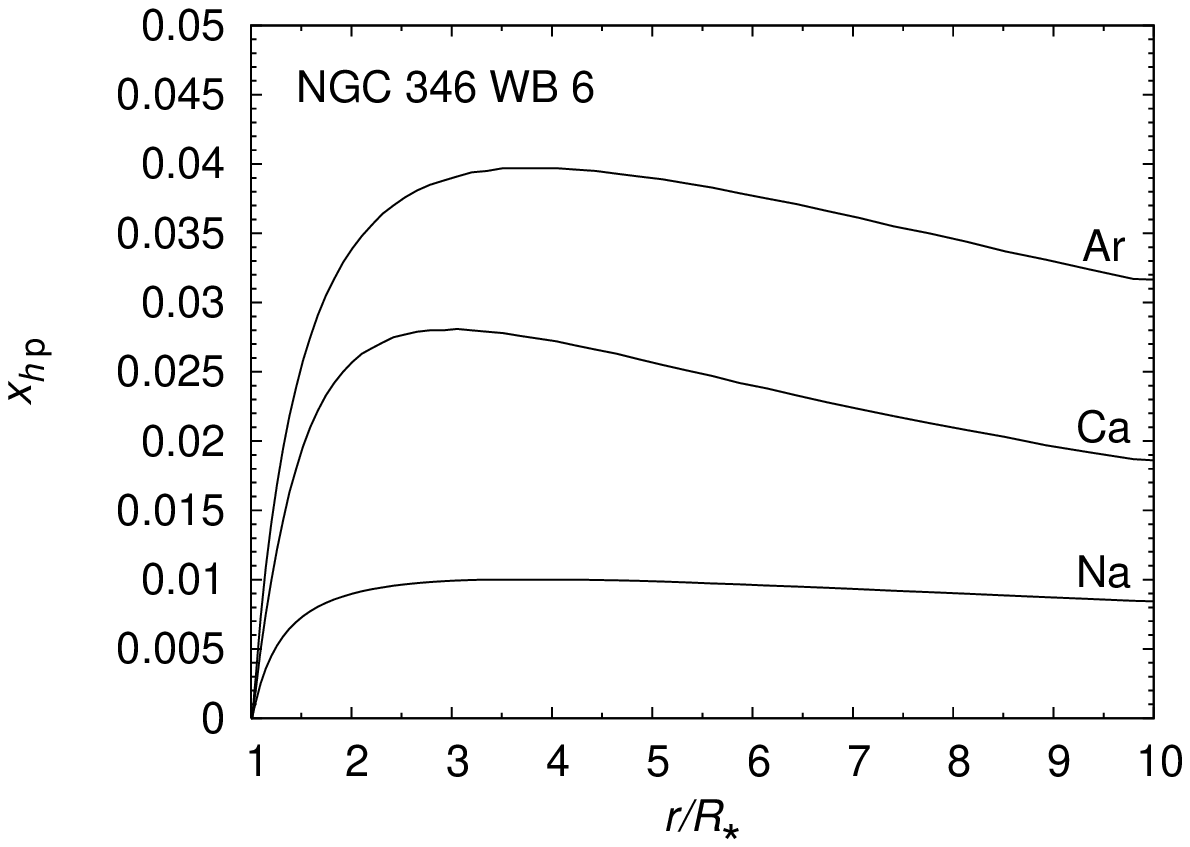}}
\resizebox{0.31\hsize}{!}{\includegraphics{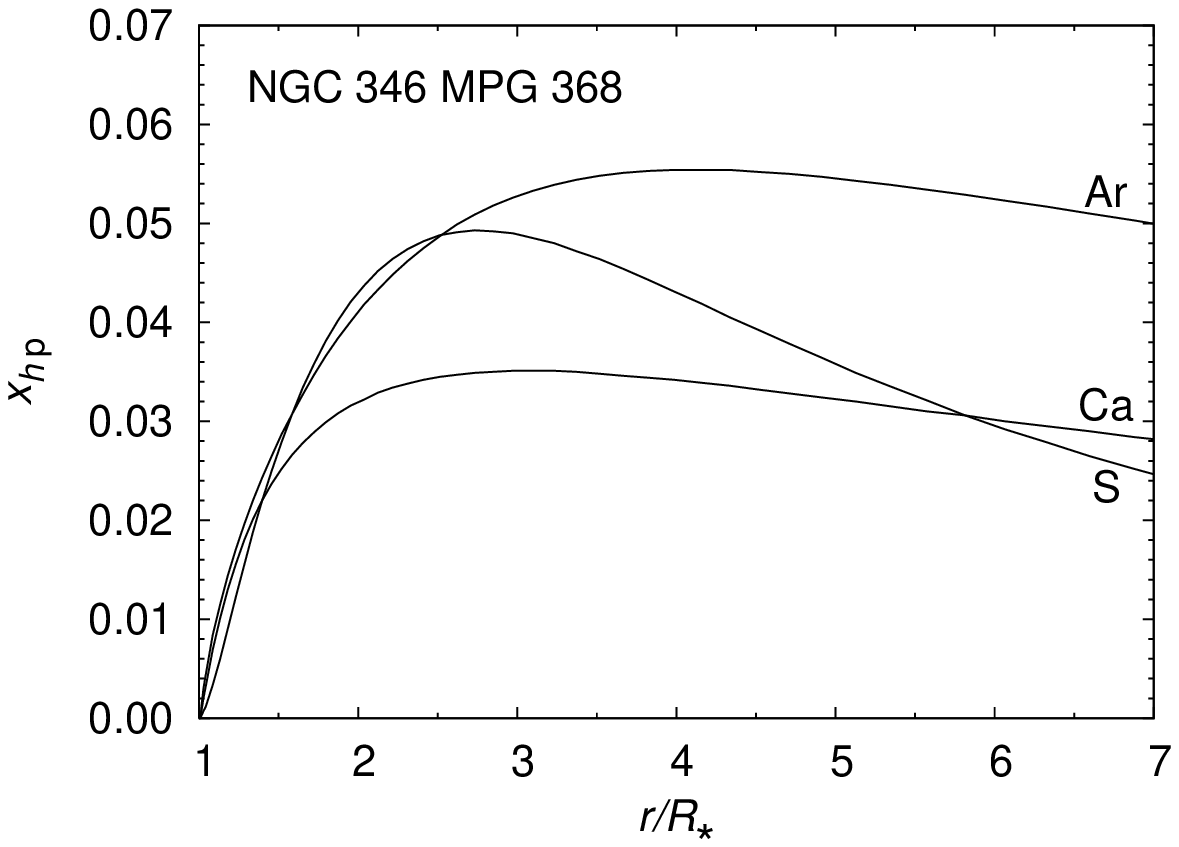}}
\resizebox{0.31\hsize}{!}{\includegraphics{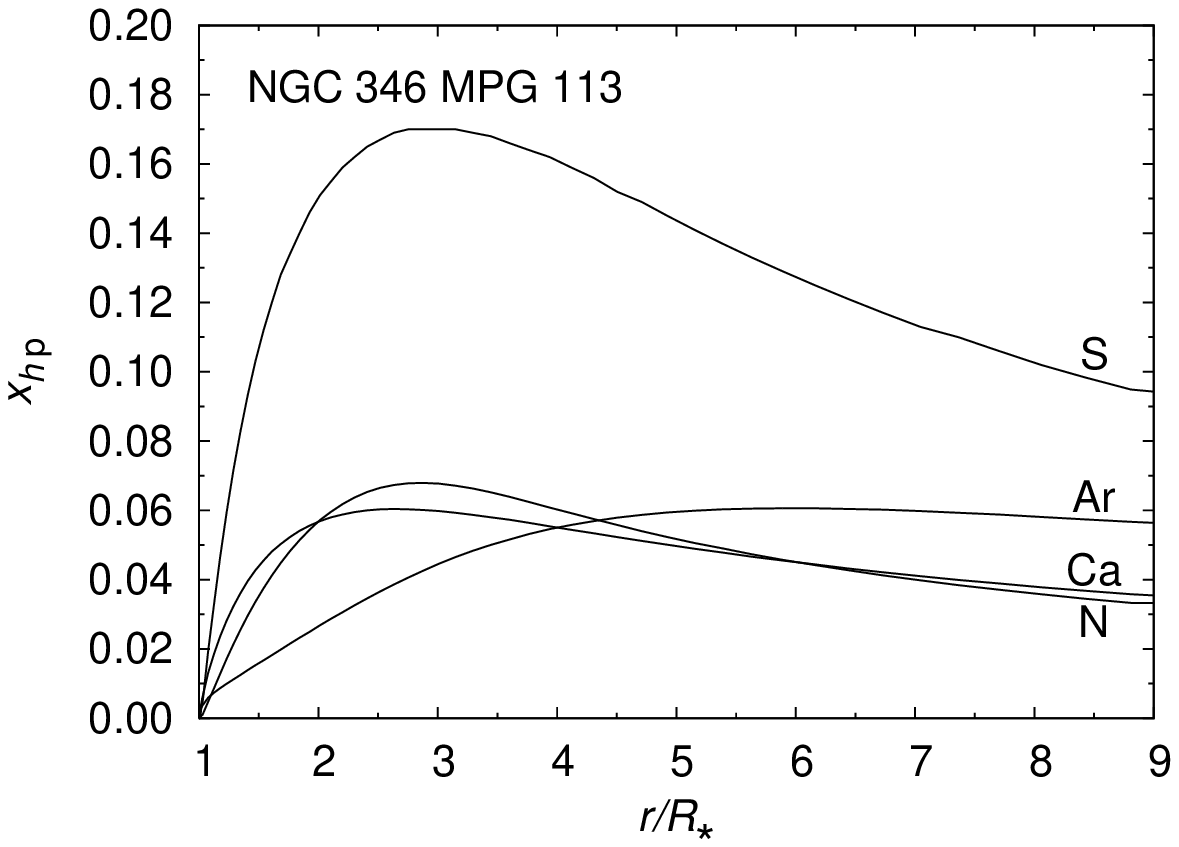}}
\resizebox{0.31\hsize}{!}{\includegraphics{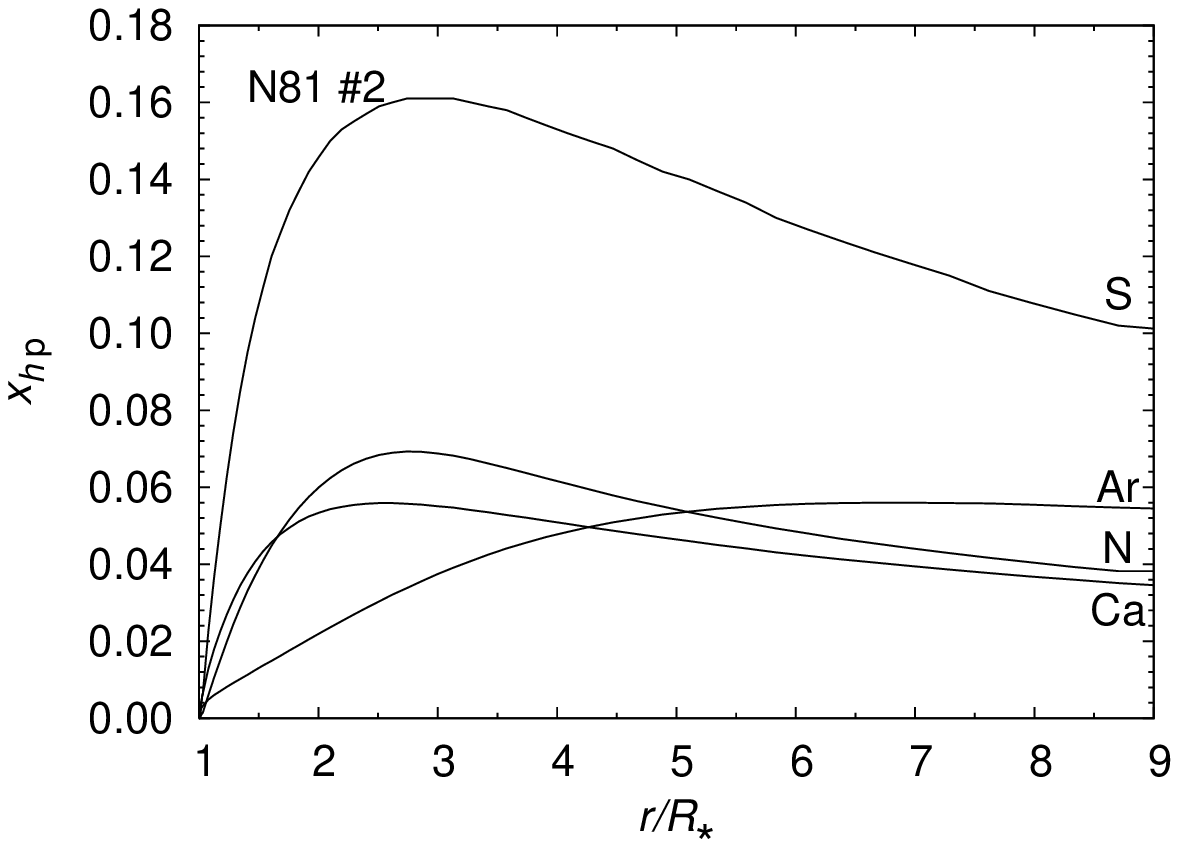}}
\resizebox{0.31\hsize}{!}{\includegraphics{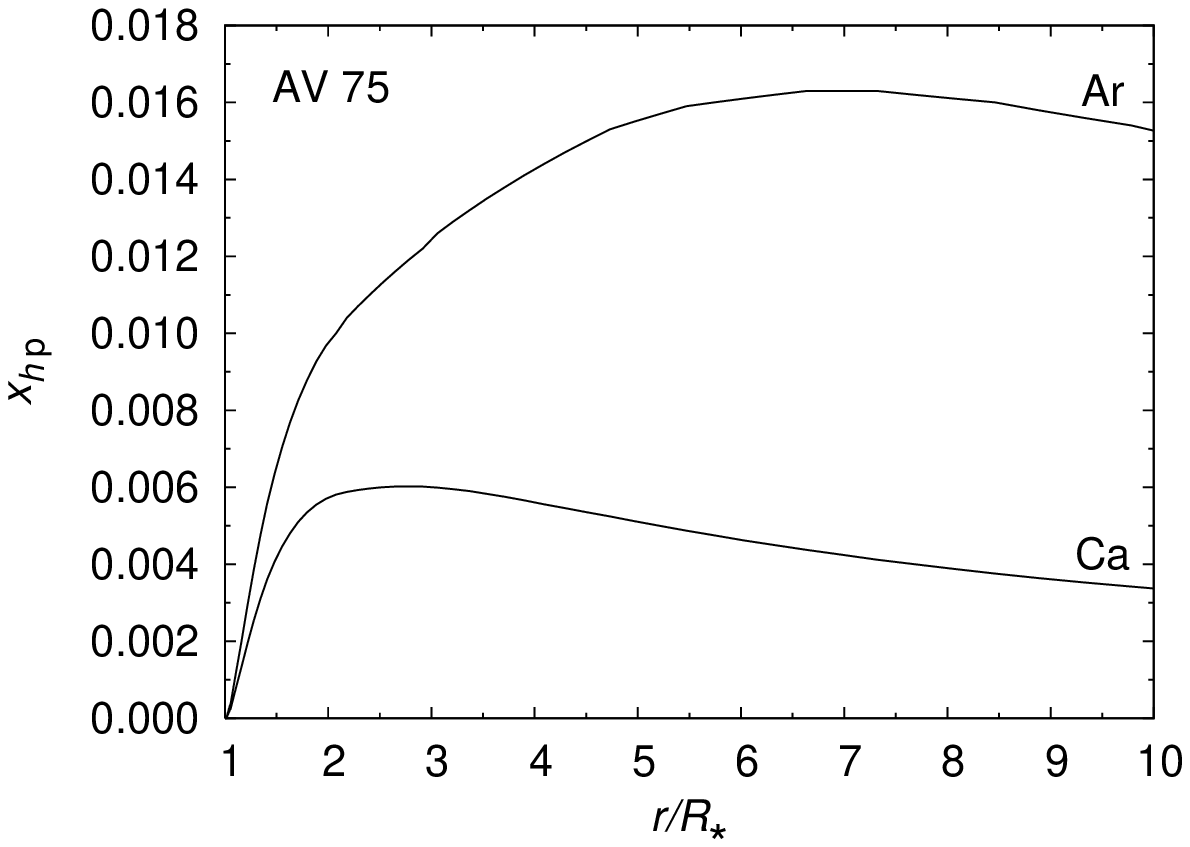}}
\resizebox{0.31\hsize}{!}{\includegraphics{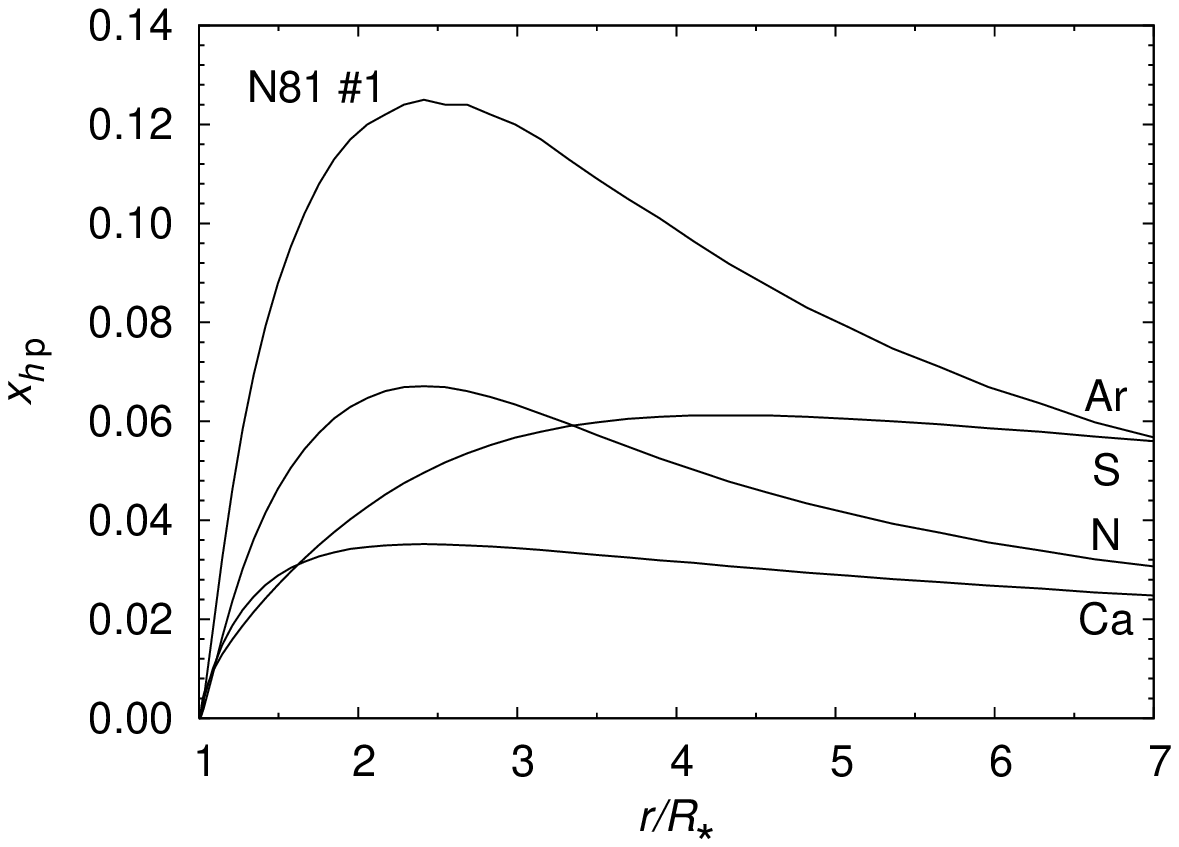}}
\resizebox{0.31\hsize}{!}{\includegraphics{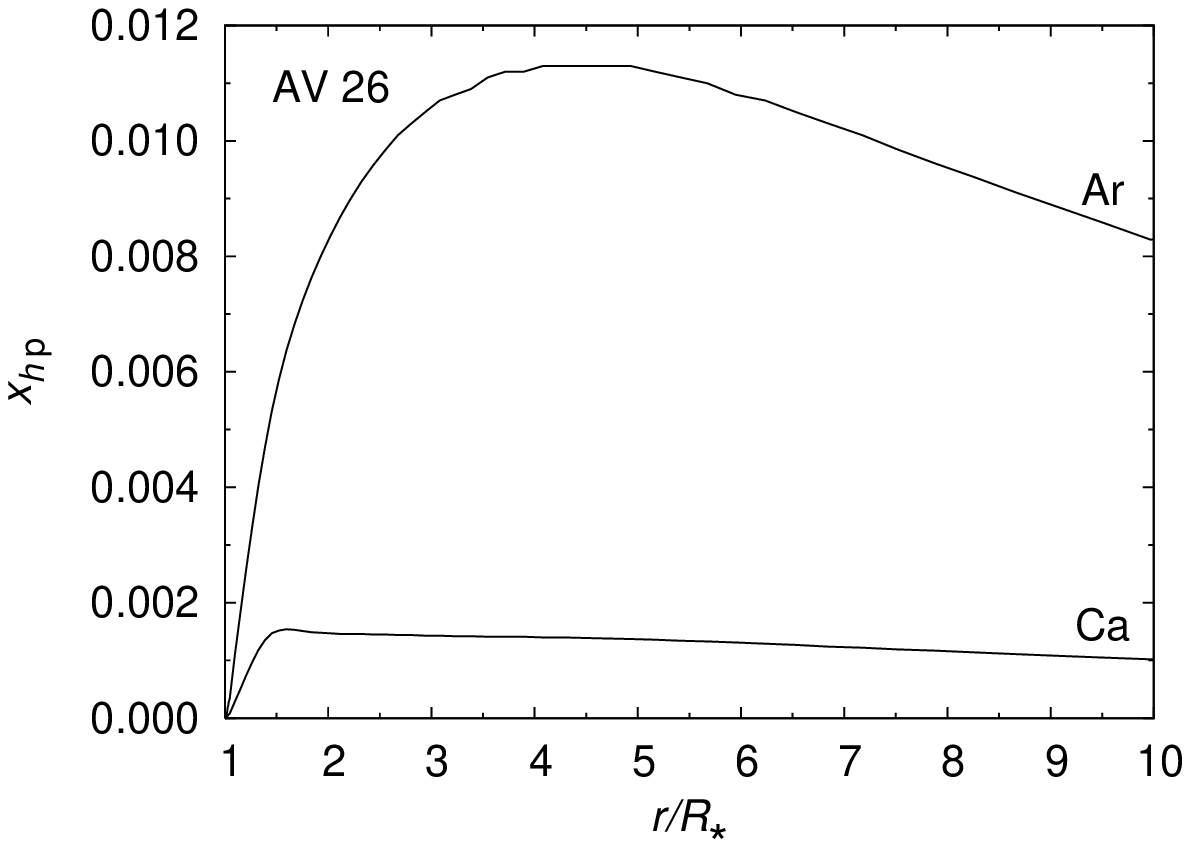}}
\resizebox{0.31\hsize}{!}{\includegraphics{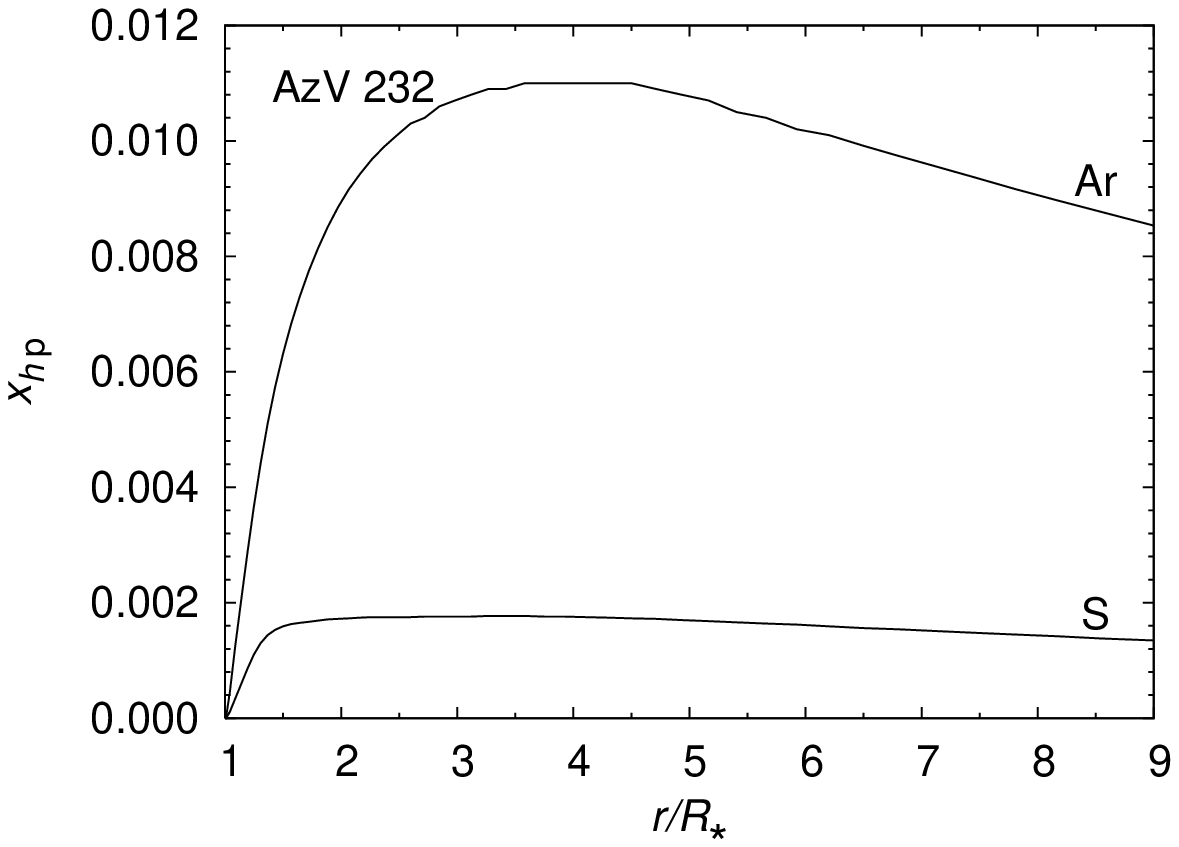}}
\resizebox{0.31\hsize}{!}{\includegraphics{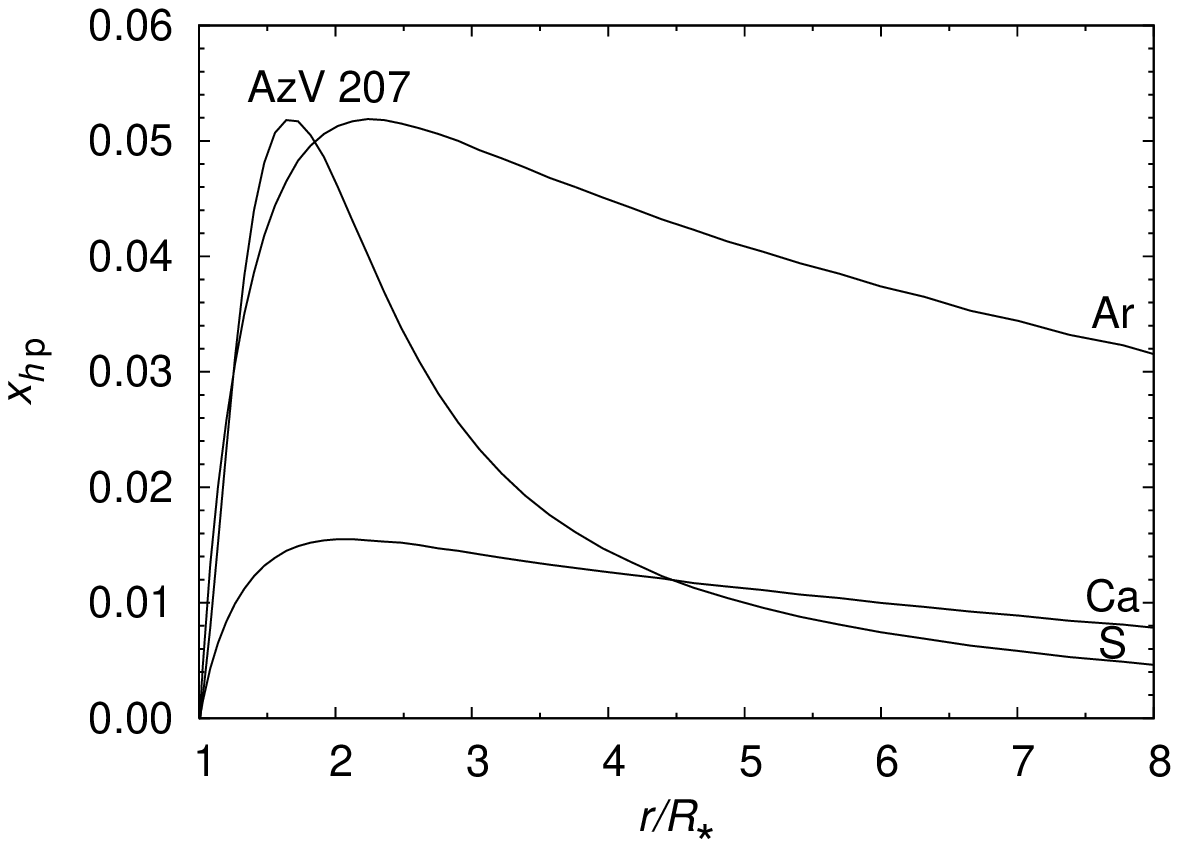}}
\resizebox{0.31\hsize}{!}{\includegraphics{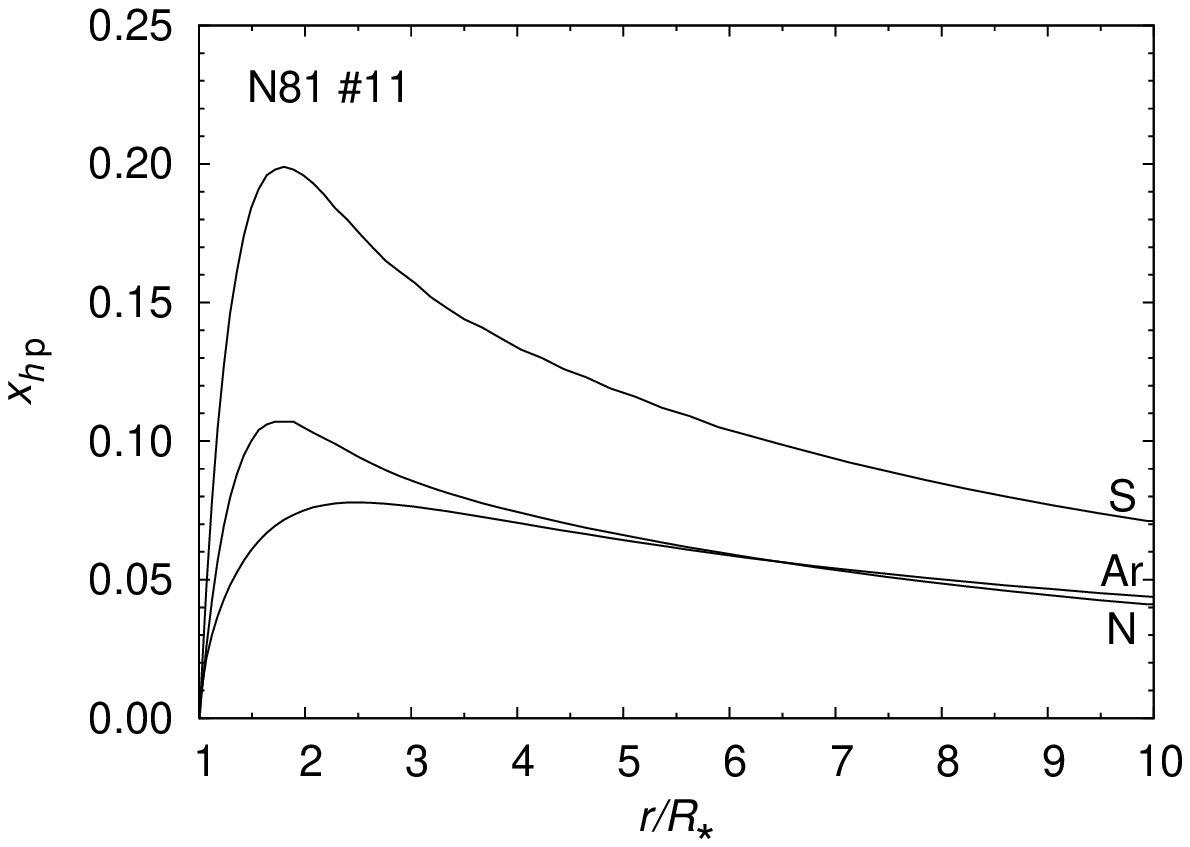}}
\resizebox{0.31\hsize}{!}{\includegraphics{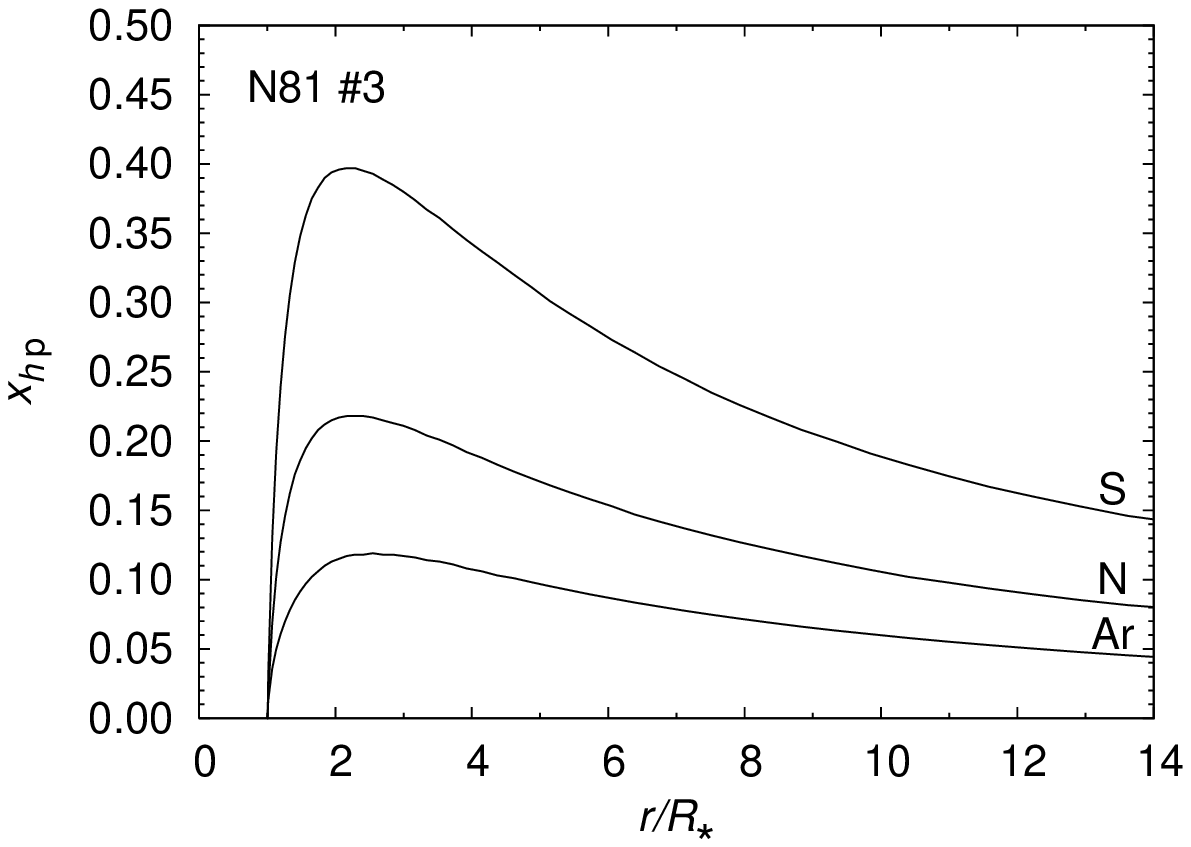}}
\resizebox{0.31\hsize}{!}{\includegraphics{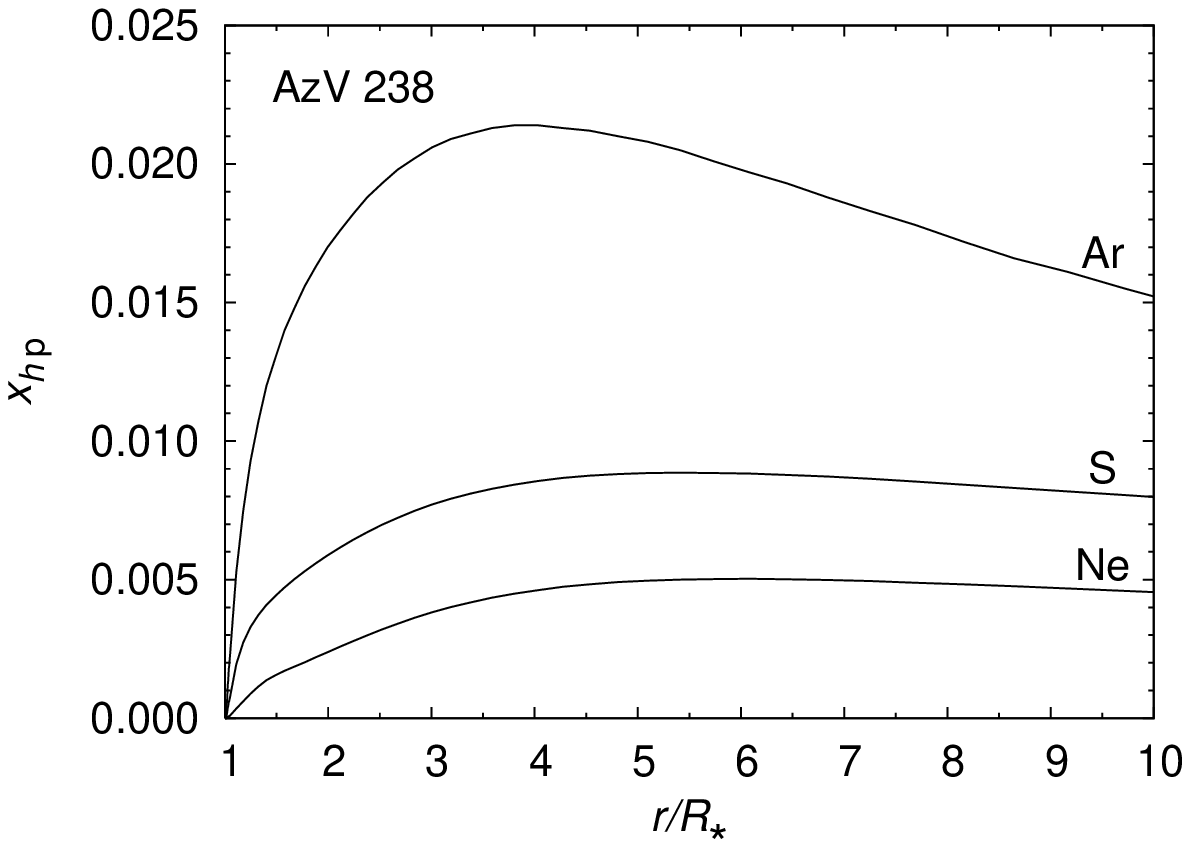}}
\resizebox{0.31\hsize}{!}{\includegraphics{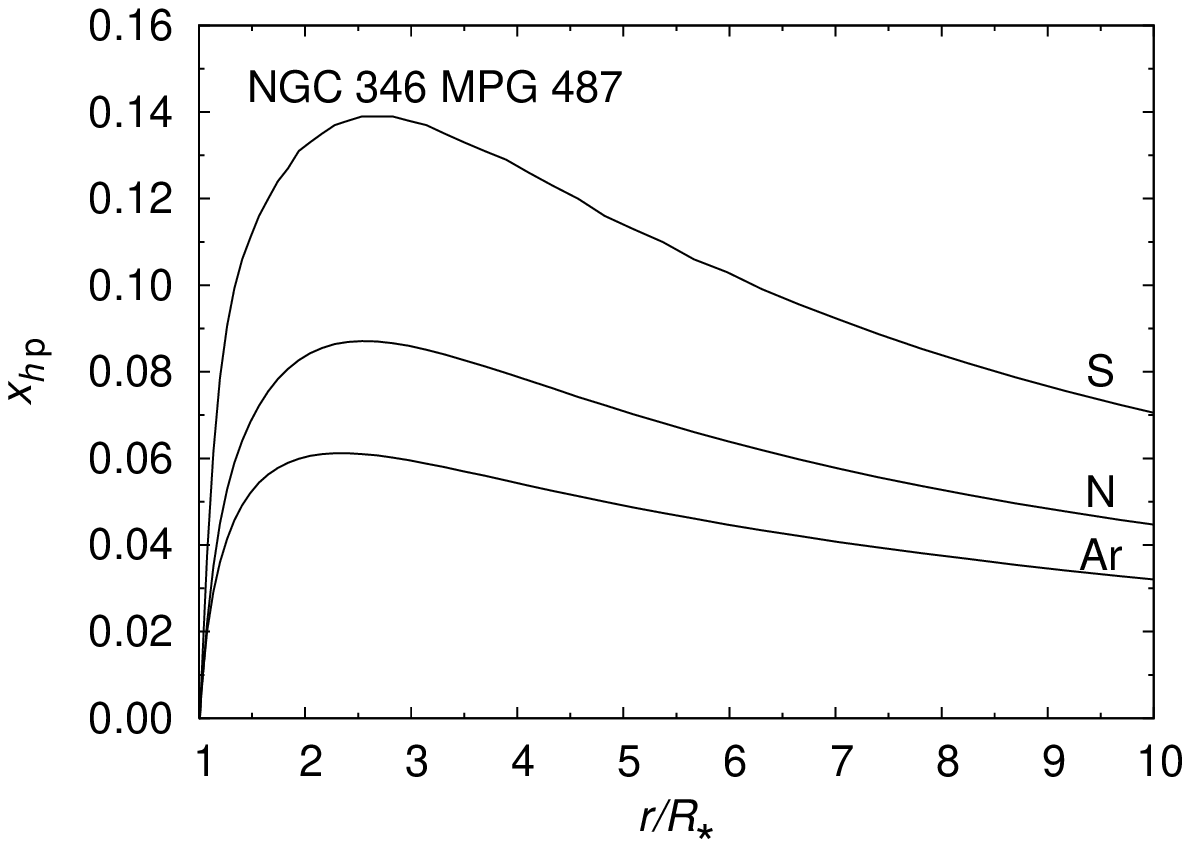}}
\resizebox{0.31\hsize}{!}{\includegraphics{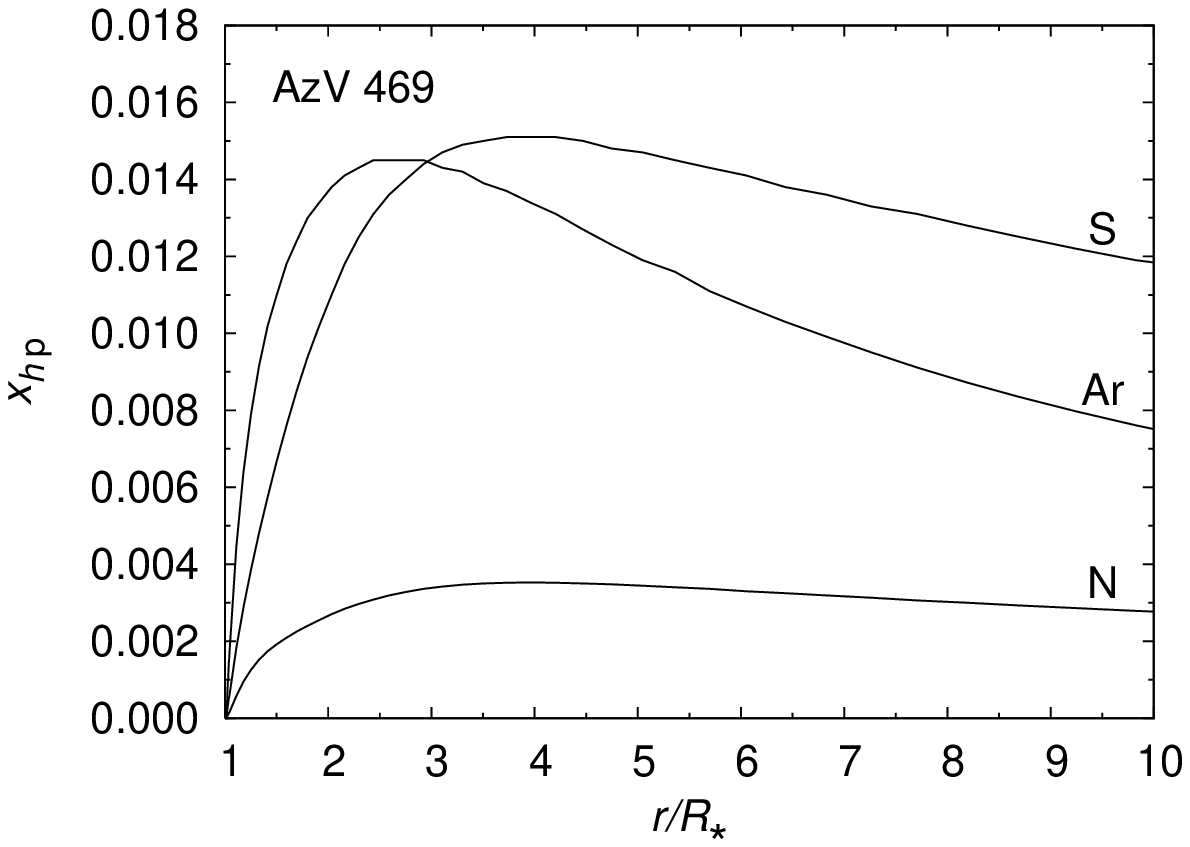}}
\resizebox{0.31\hsize}{!}{\includegraphics{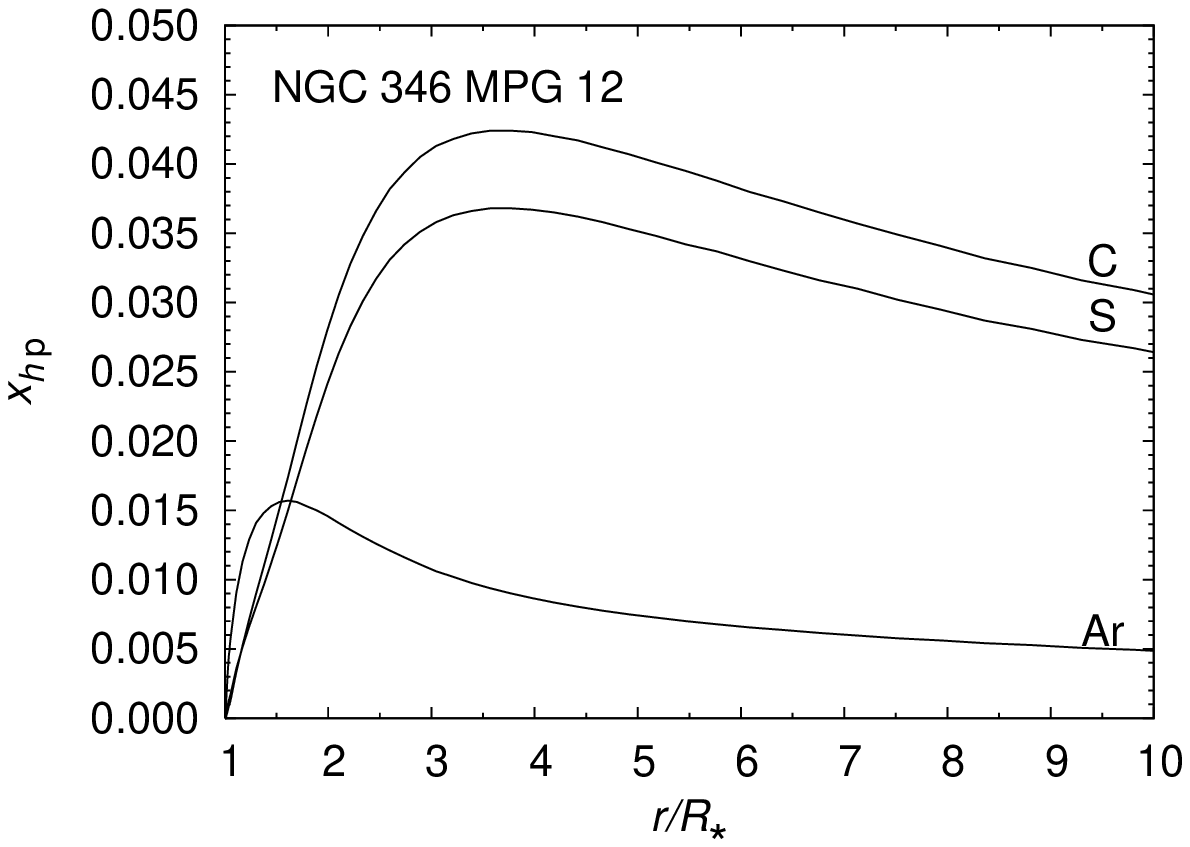}}
\caption[]{Relative velocity differences between individual {heavier} elements
and passive component calculated using Eq.~\eqref{berlin}. {Only values for
elements with highest relative velocity differences are plotted. Velocity
differences for other heavier elements are much lower.}}
\label{xpiobr} 
\end{figure*}

\begin{figure}
\resizebox{\hsize}{!}{\includegraphics{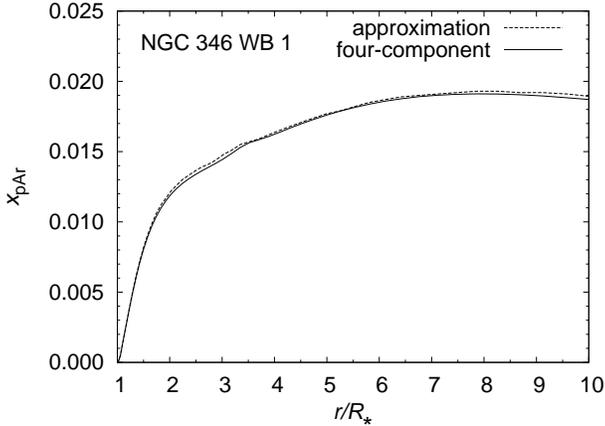}}
\caption[]{Comparison of relative velocity differences between passive
component and argon calculated using
approximate expression Eq.(\ref{berlin}) and using four-component model.
{Clearly, in the case of negligible frictional heating  Eq.(\ref{berlin}) is
able to correctly predict velocity differences.}}
\label{chocen}
\end{figure}

\begin{figure*}
\centering
\resizebox{0.45\hsize}{!}{\includegraphics{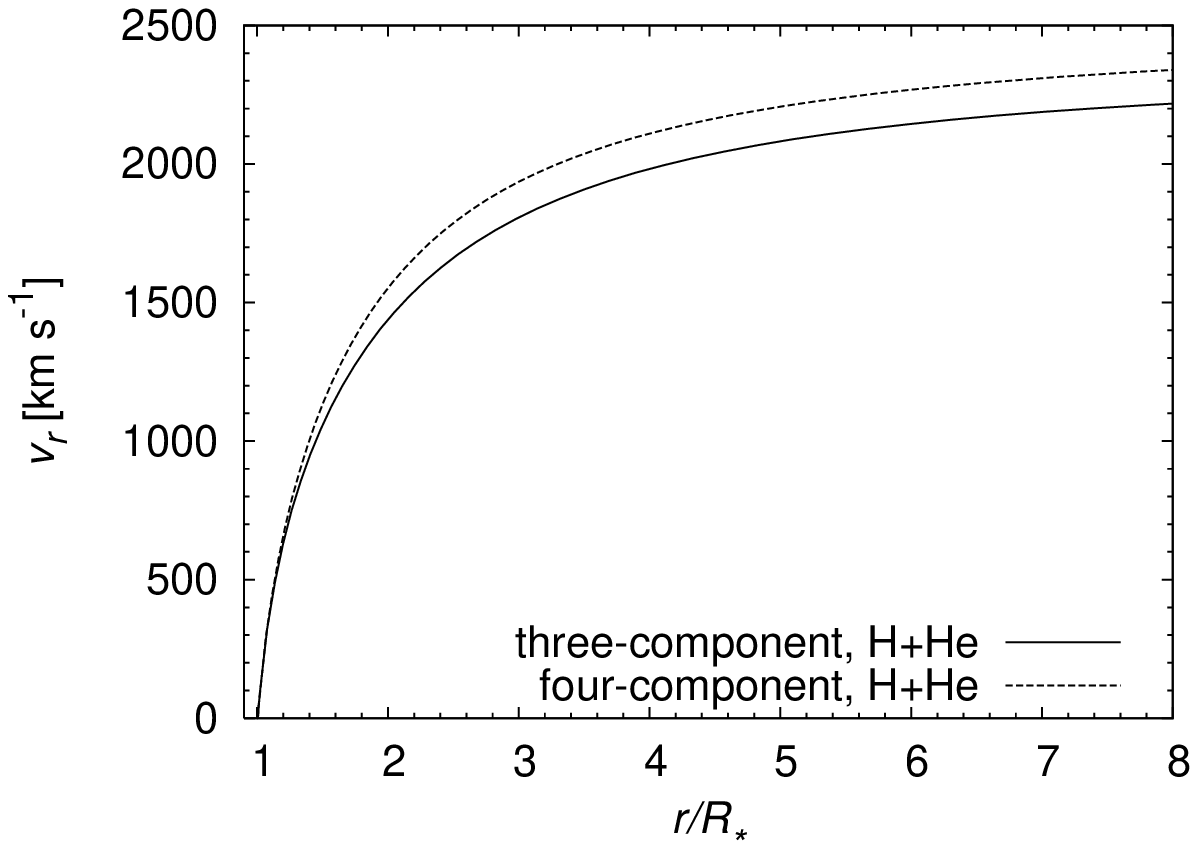}}
\resizebox{0.45\hsize}{!}{\includegraphics{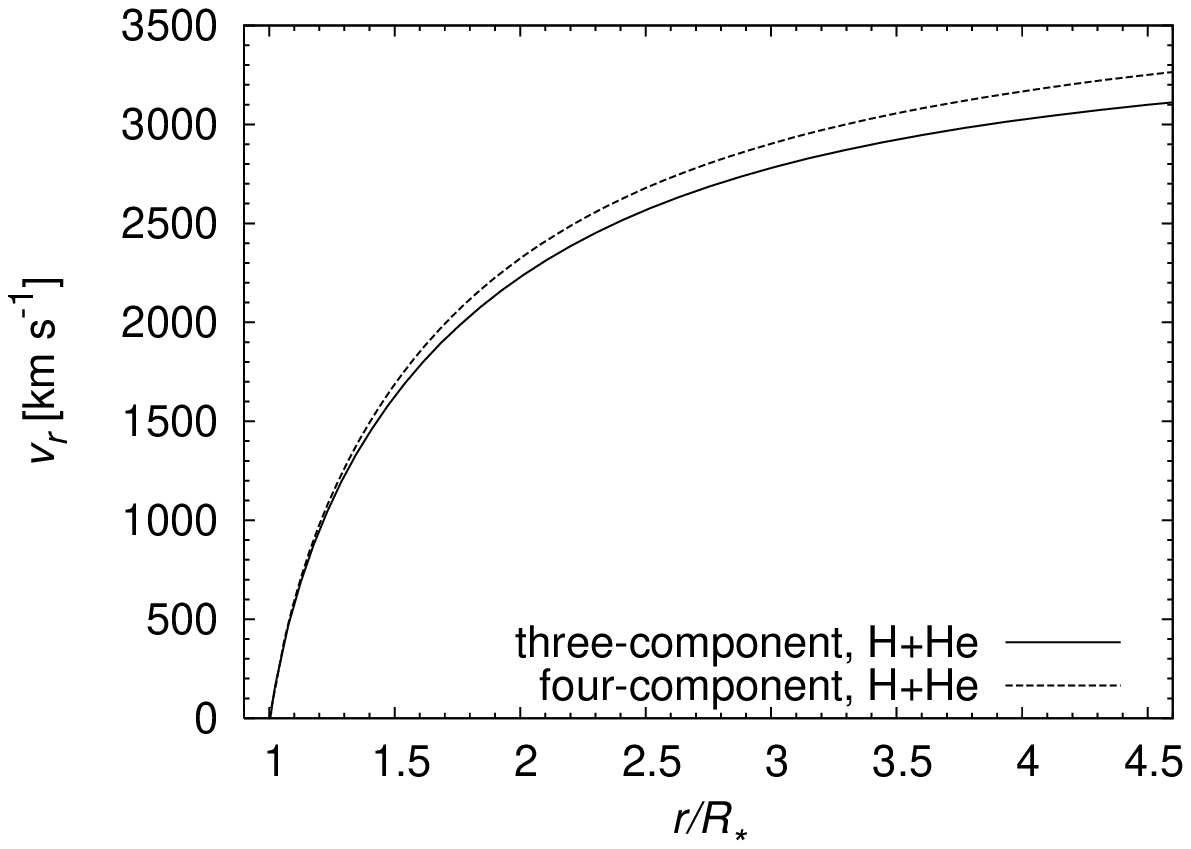}}
\resizebox{0.45\hsize}{!}{\includegraphics{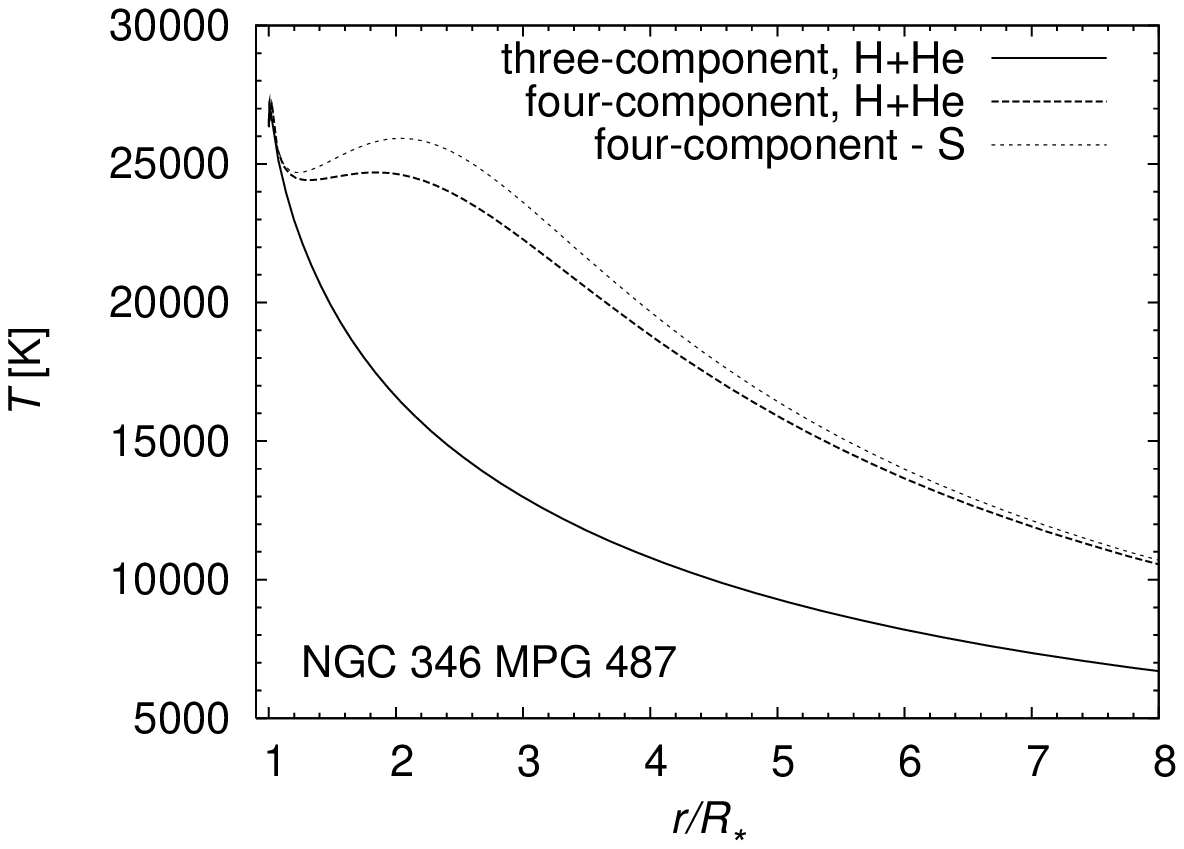}}
\resizebox{0.45\hsize}{!}{\includegraphics{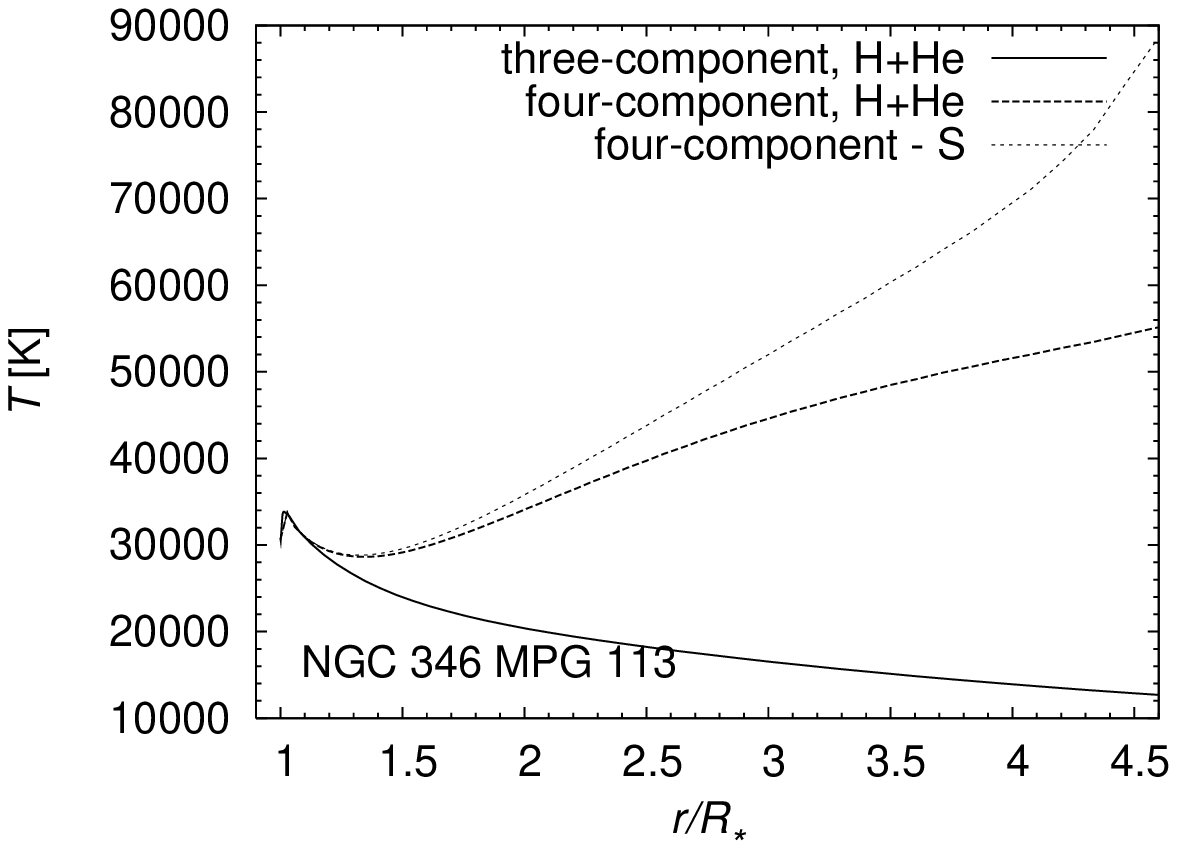}}
\resizebox{0.45\hsize}{!}{\includegraphics{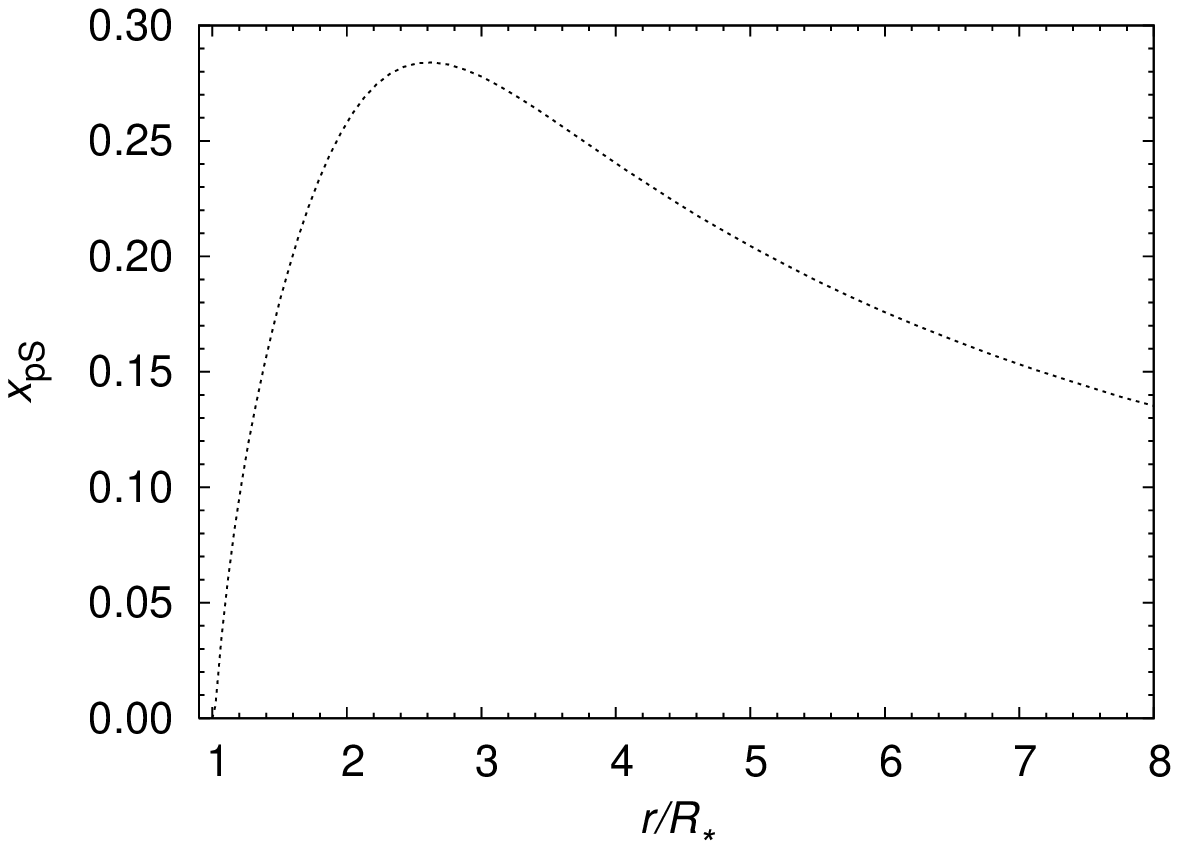}}
\resizebox{0.45\hsize}{!}{\includegraphics{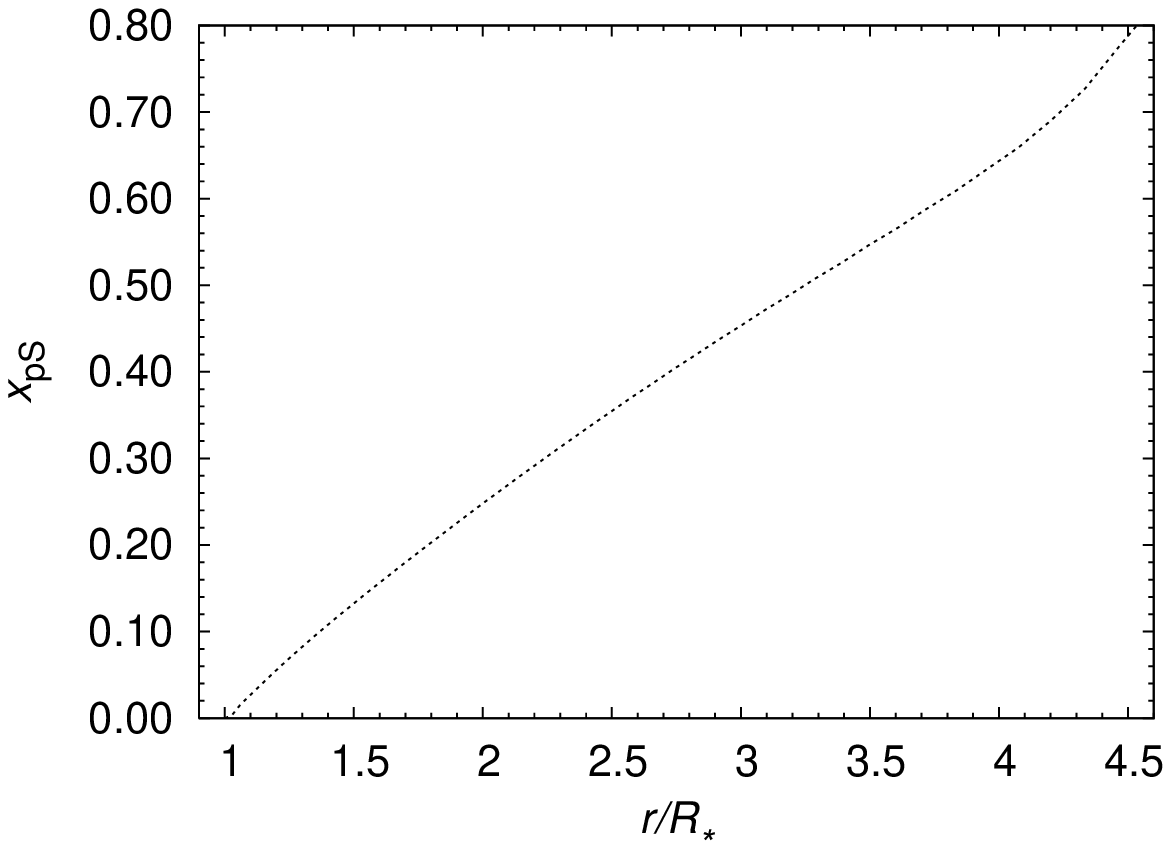}}
\caption[]{Comparison of four-component model (with sulfur as a fourth component,
dashed line) and three-component model (solid line) for star NGC 346 MPG 487
(left) and
NGC 346 MPG 113 (right). {\em Top panel}: Comparison
of velocity of passive component (hydrogen+helium) in individual models. Note
that the velocities of \zm{all} individual wind components in four-component or
three-component models are similar. {\em
Middle  panel}: Comparison of wind temperatures. Wind temperatures of individual
wind components in three-component model are nearly equal. Wind temperature
in four-component model is higher due to the frictional heating and sulfur
temperature is even higher than the temperature of other components. {\em
Bottom panel}: Relative velocity difference between passive component and
sulphur in four-component model.
{The velocity difference in the case of NGC 346 MPG 113 is close to one in
the outer wind regions, thus potentially enabling {instability due to} sulphur decoupling.}}
\label{doubravice}
\end{figure*}

\begin{figure*}
\centering
\resizebox{0.45\hsize}{!}{\includegraphics{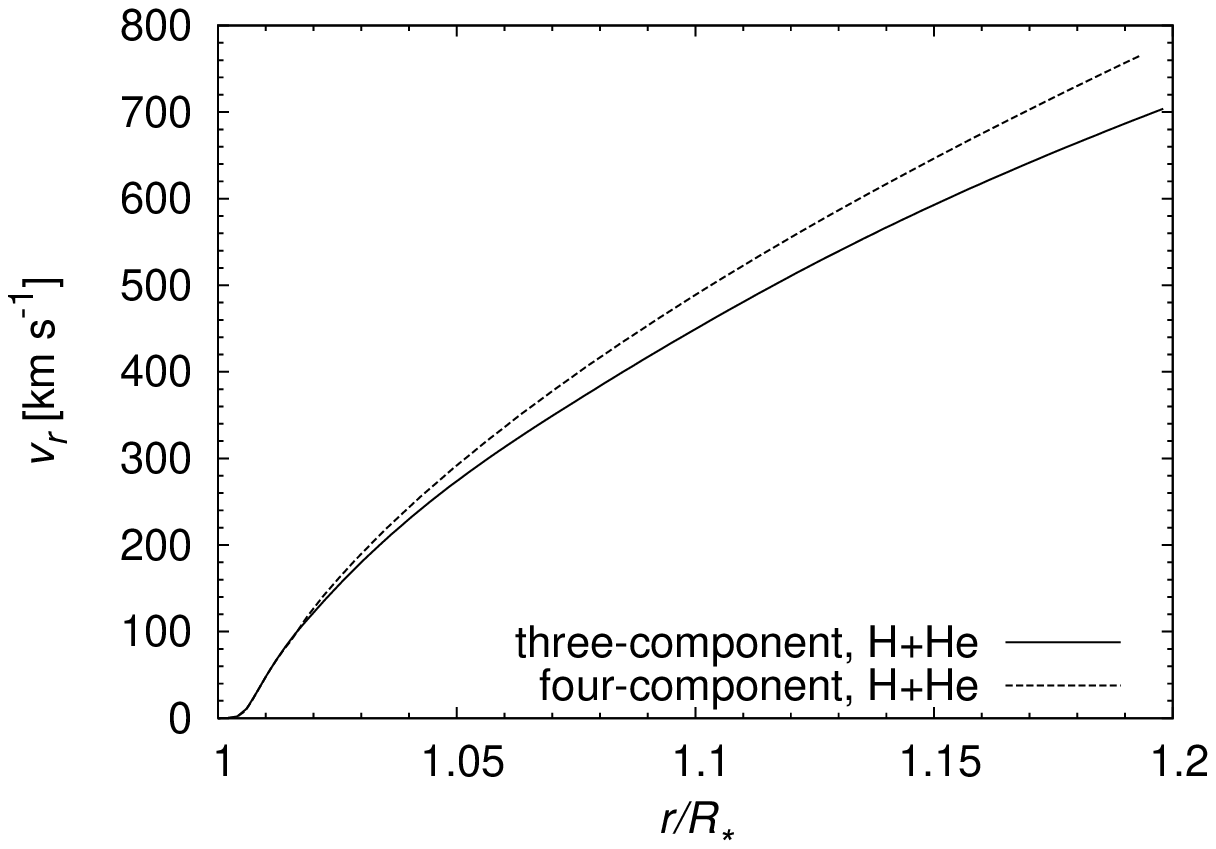}}
\resizebox{0.45\hsize}{!}{\includegraphics{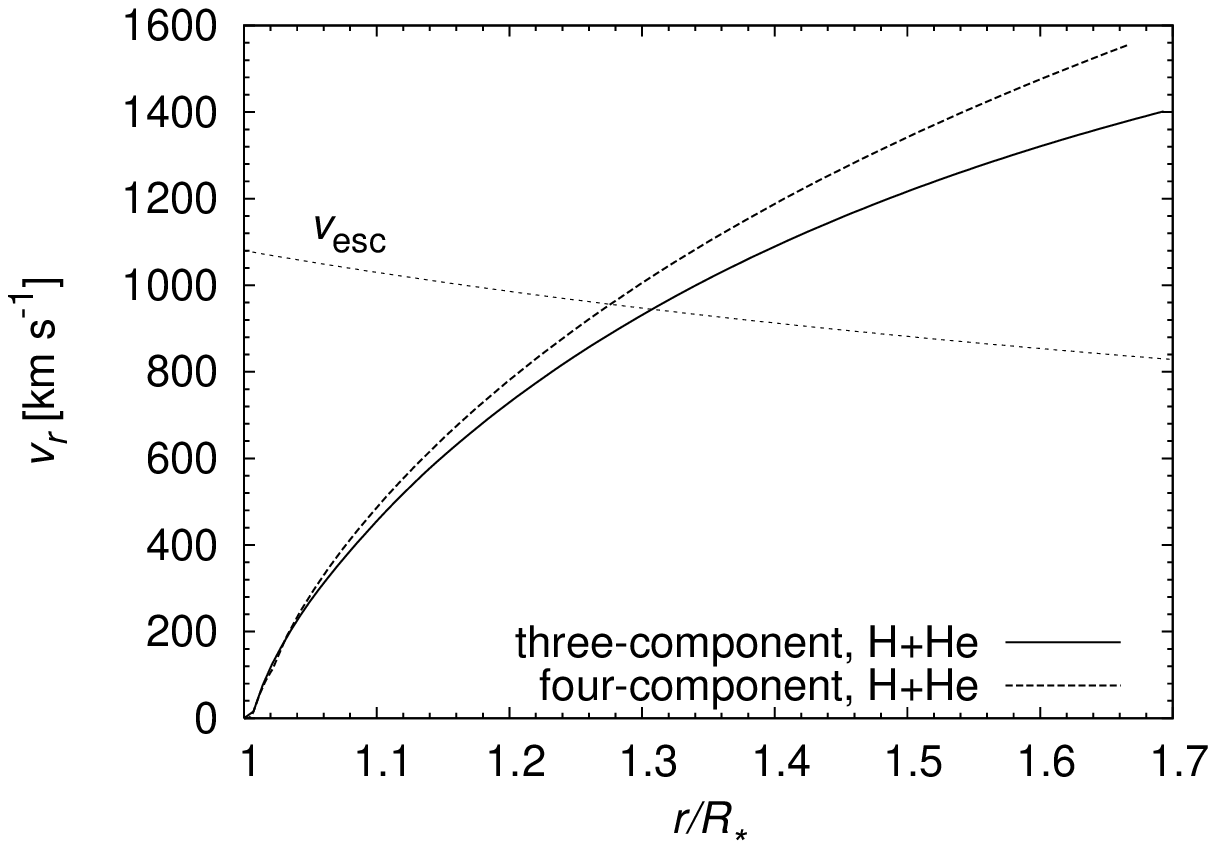}}
\resizebox{0.45\hsize}{!}{\includegraphics{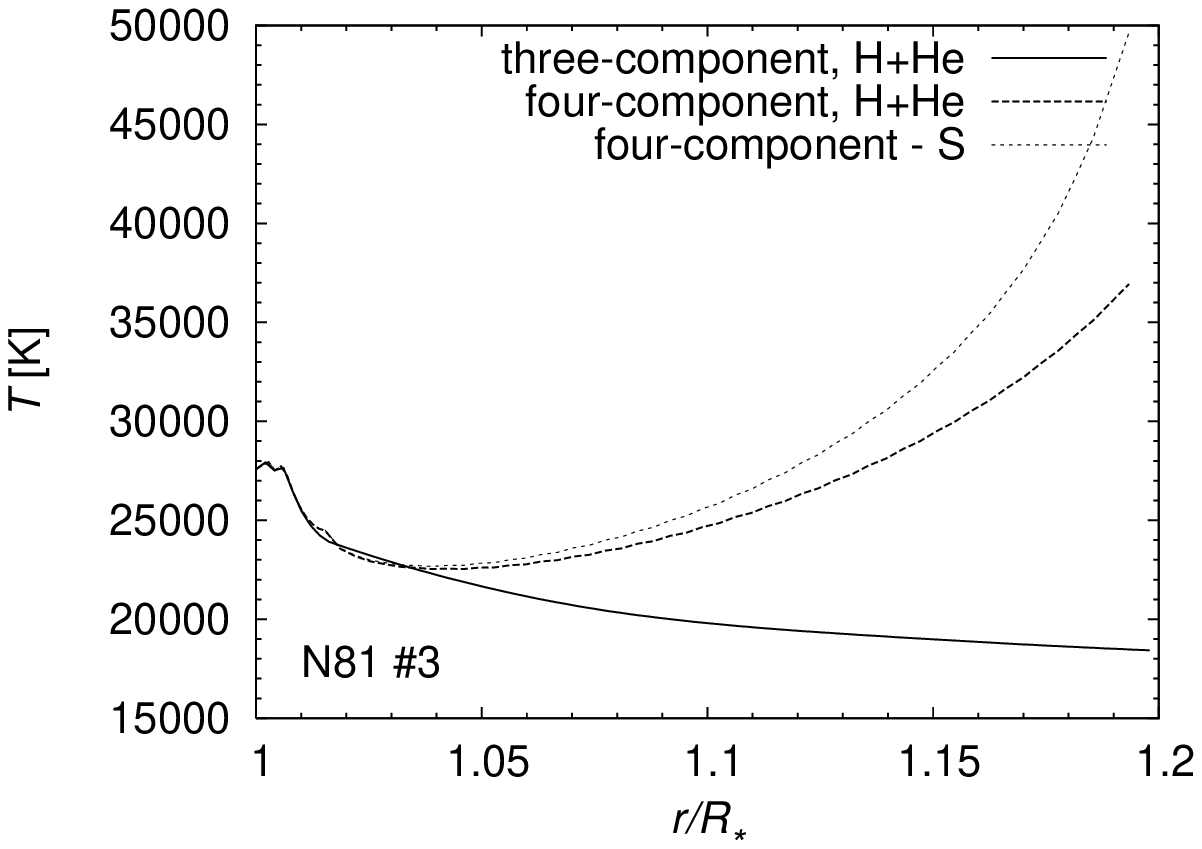}}
\resizebox{0.45\hsize}{!}{\includegraphics{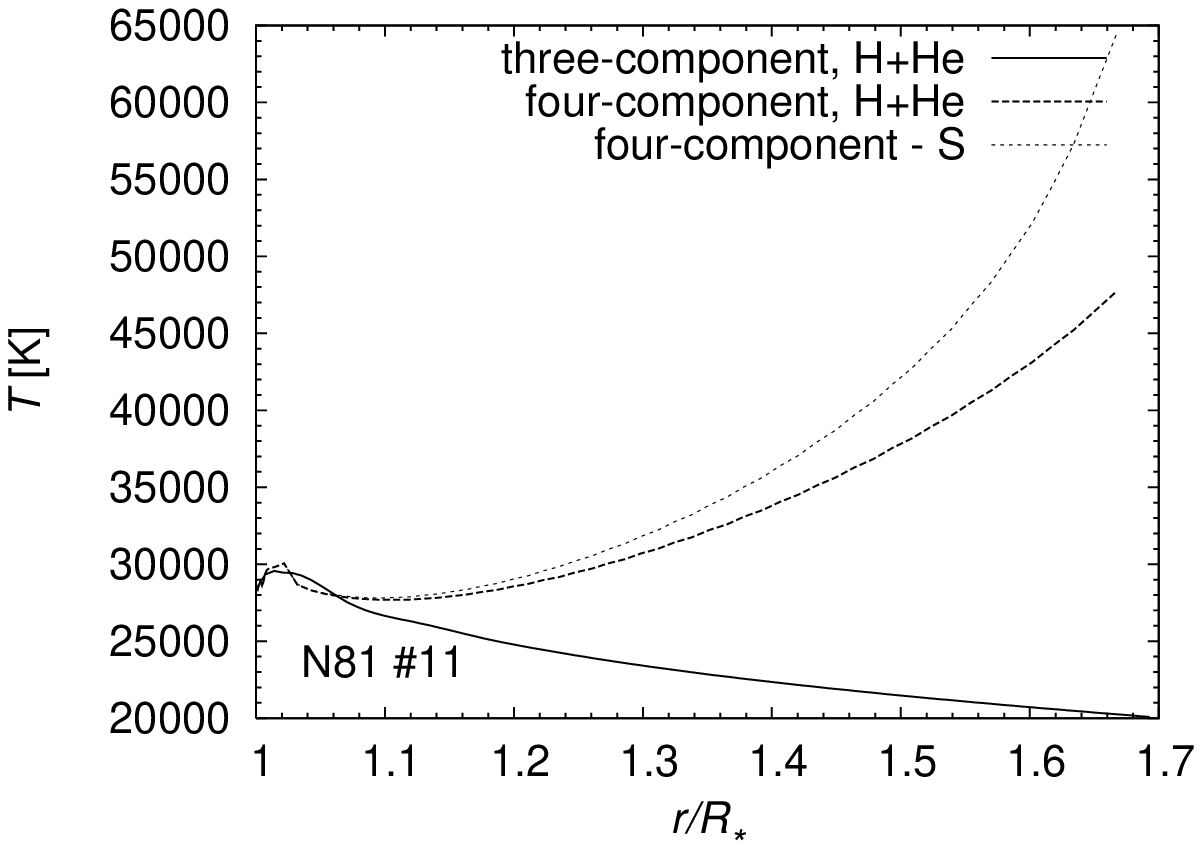}}
\resizebox{0.45\hsize}{!}{\includegraphics{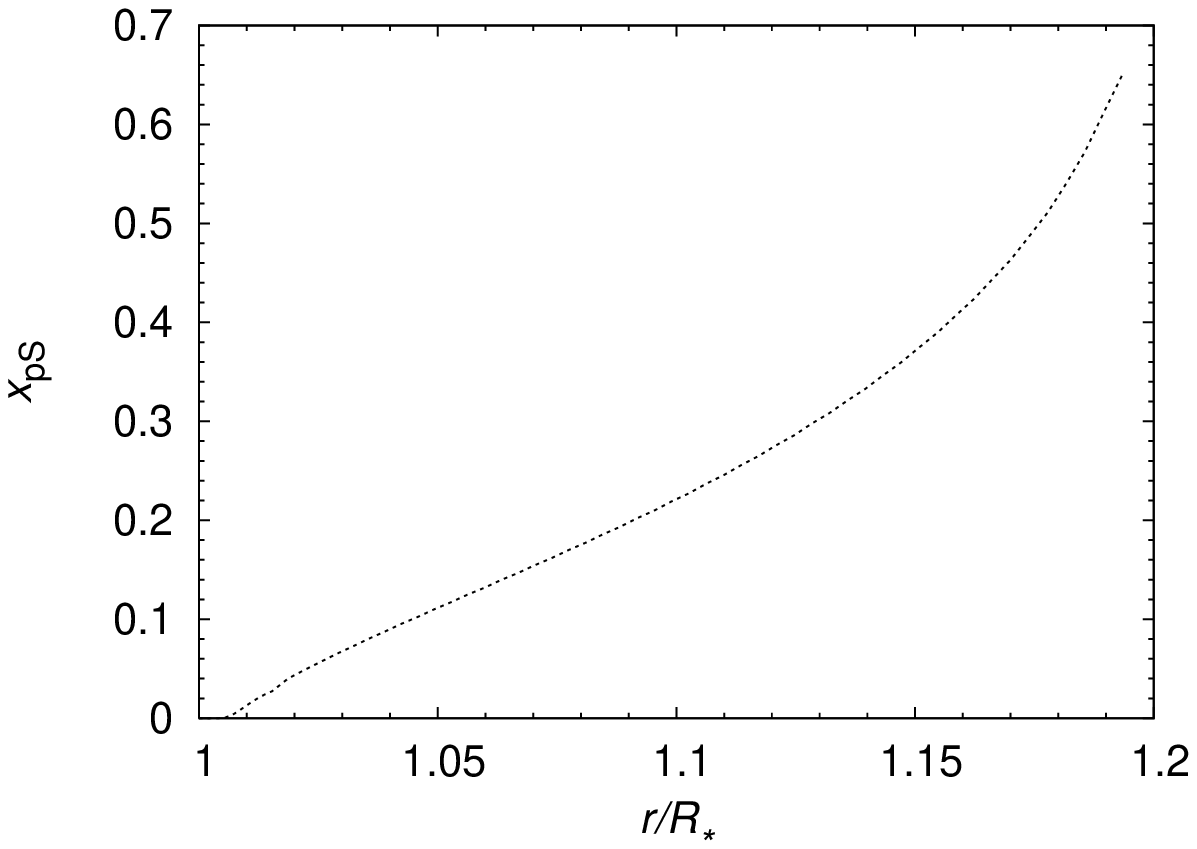}}
\resizebox{0.45\hsize}{!}{\includegraphics{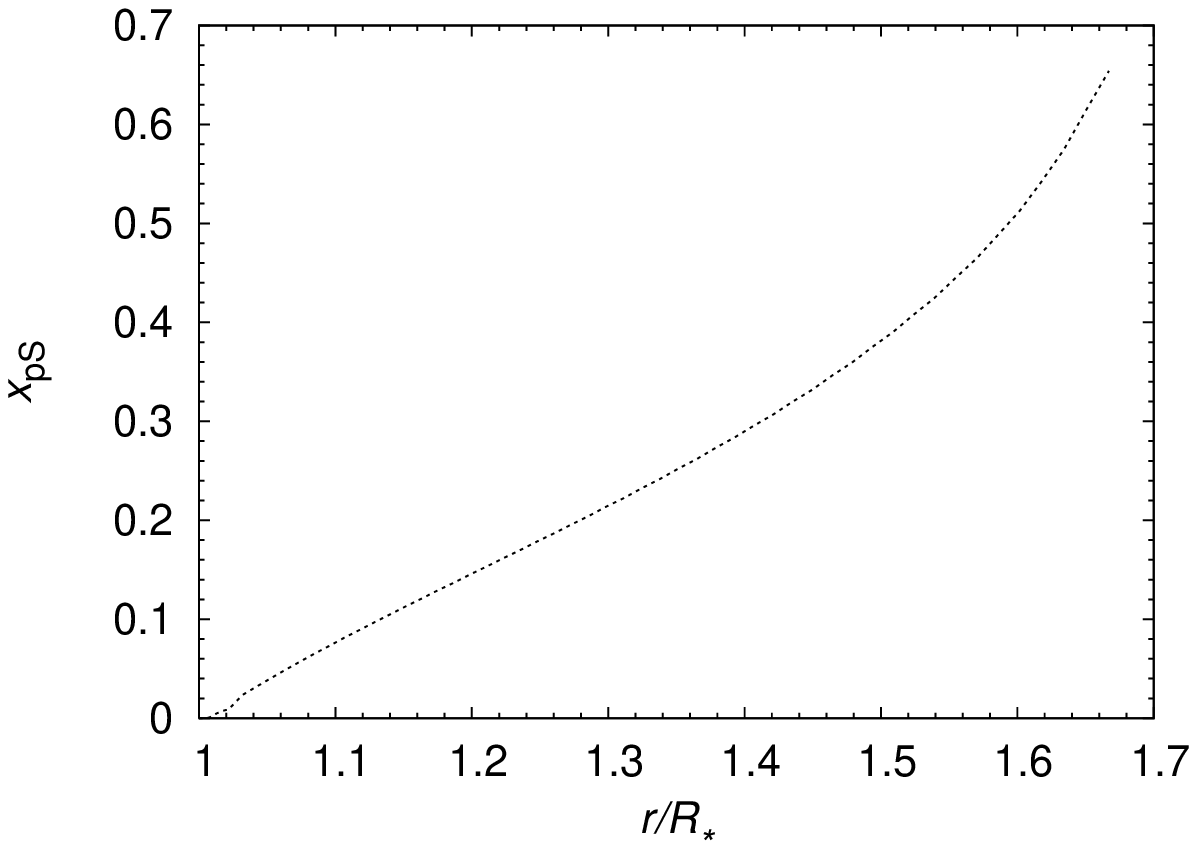}}
\caption[]{{The same as Fig.~\ref{doubravice} for SMC-N81 stars \#3 ({\em
left}) and
\#11 ({\em right}). Note that for winds of both stars the friction between sulphur and passive
wind component (hydrogen and helium) is important for wind structure. In both
cases the sulphur decoupling {may} occur. In the case of SMC-N81 \#3 sulphur
{may} decouple for velocities lower than the escape speed {(the surface escape
speed for this star is $v_\text{esc}=1180\kms$)} and in the case of SMC-N81
\#11 sulphur {may} decouple for velocities slightly higher than the escape speed
(the escape speed is also plotted on the graph {of this star}).}}
\label{reckovice}
\end{figure*}

To {test the possibility of frictional heating}, we have calculated four-component wind
models, where the fourth component is either argon, sulphur or carbon, depending
on which element for a given star has a maximum velocity difference. The other
three components are {heavier} ions, passive component (hydrogen and helium)
and {free} electrons.

First, we tested whether the simple equation
Eq.~(\ref{berlin}) is able to reliably predict the velocity
differences. We plot the result only for the case of star NGC~346~WB~1
(see~Fig.\ref{chocen}), however the result {that} Eq.~(\ref{berlin}) is able
to very reliably predict the approximate velocity differences {is} similar also for other stars.

For stars with highest velocity differences the frictional heating
occurs\footnote{%
{In the common case that occurs in the stellar atmosphere and wind it is
usually not necessary to differentiate between kinetic temperatures of individual
particles since these temperatures are nearly equal. However, in the present
case the temperatures of individual wind components significantly differ. Since
the collisional terms that enter into the statistical equilibrium equations are
caused by the collisions with free electrons that move with much higher
thermal speed than other particles, we inserted the electron temperature into all terms in the
statistical equilibrium equations. This is indeed only 
approximative in the case of relatively large velocity differences between wind
components.}}.
This is shown in Figs.~\ref{doubravice}, {\ref{reckovice}}, where we compare four-component models
(with wind components sulphur, {remaining}
{heavier} ions, passive component and electrons)
with three-component models (with wind components {heavier} ions, passive
component and electrons). In the outer wind regions the relative
velocity difference between passive component and sulphur $x_\text{pS}$ is close
to $1$. Due to this higher velocity difference the stellar wind is frictionally
heated in the outer parts. Consequently, wind temperature is higher in
four-component model than in three-component one. Higher wind 
temperature also slightly modifies \zm{the} radiative force since wind velocity is
slightly higher in four-component case.

\zm{The frictional heating also slightly
modifies the ionization equilibrium. The ionization fraction of higher
ionization stages (such as \ion{N}{v}, \ion{O}{v}) of frictionally heated wind
is higher than in the wind with negligible frictional heating, whereas the
ionization fractions of lower ionization stages (e.g.~\ion{C}{iv}, \ion{Si}{iv})
are lower in the frictionally heated wind. The differencies are by about a
factor of~$2$.}

{The velocity differences in the case of frictionally heated four-component
wind models are larger than those estimated using approximate Eq.~\eqref{berlin}
(cf.~Fig.~\ref{xpiobr}), since velocity difference increases with increasing
temperature. The velocity difference in the case of NGC 346 MPG 113{,
SMC-N81 \#11 and \#3} is so high that the wind {instability connected with
sulphur decoupling may occur (Owocki \& Puls \citeyear{op},
Krti\v{c}ka \& Kub\' at \citeyear{kkiii})}.
{For star \#3 this occurs for velocities lower than the
escape speed.}

We have shown that for stars {that} have very low observed mass-loss rates the
velocity differences {for} some {heavier} elements are high. {In such
case}
either frictional heating becomes important or even some elements may decouple
from the wind. Note that these multicomponent effects are important
already for
theoretically derived mass-loss rates, i.e. those that are much higher than the
observed ones.} \zm{Since the frictional heating increases the wind
temperature only in the outer wind regions (for velocities higher than that
corresponding to the critical point velocity), its influence on the mass-loss
rate is negligible.}

Fall back of the wind material (which may cause the lowering of wind
mass-loss rate) may occur only if the wind components decouple 
below the point where wind velocity is equal to the escape velocity.
{Only in this case
the runaway instability connected with decoupling 
may lead to fall back of wind material
(Porter \& Skouza \citeyear{reac}, Votruba et al. \citeyear{ufosap}) and
consequently lower the mass-loss rate.
This occurs only for star SMC-N81 \#3. For other stars}
we have not obtained the
possibility of decoupling close to the stellar surface, for velocities lower
than the escape velocity.  This means that presented models are, {with one
exception}, close to the star stable for perturbations of multicomponent flow. 
{Consequently, we are not able to explain very low
observed mass-loss rates using multicomponent models for all except one star.
Only for star SMC-N81 \#3
the low observed mass-loss rate may be connected with multicomponent effects.}


\section{Discussion and conclusions}

We have presented NLTE wind models {of cooler} O stars in SMC. These
models enable prediction of basic wind parameters, i.e. {the} mass-loss rates
and {the} 
terminal velocities. We have compared these predicted wind parameters with those
derived from observation. We have concluded that there is a relatively good
agreement between observed and predicted terminal velocities with some scatter.
This scatter can be partly attributed to the poorly known chemical
composition of SMC stars. Moreover, the existence of unidentified binaries in
our sample can also influence the comparison. Last but not least, the
simplification involved in our code (e.g.~{simplified treatment of radiative
transfer equation}\zm{, like neglect of line overlaps})
can also influence the results. {However, t}he situation with mass-loss rates
is different. For relatively high mass-loss rates
(i.e.~$\dmdt\gtrsim10^{-7}\,\smrok$)
there is again a good agreement between predicted quantities and those
derived from observations. However, for smaller mass-loss rates the agreement is
rather poor, predicted mass-loss rates are significantly higher than those
derived from observation. Note that this is not an effect of our models only,
since there is {a} similar disagreement between observations and models of
{VKL}  (see B03 and {Mr04}).

We have tested whether the mentioned \zm{systematic} disagreement {between theoretical and
observ{ed} values of mass-loss rates} can be caused by the
decoupling of some metals from the mean flow. We have shown that for stars
with relatively good agreement between observed and predicted mass-loss rates
the velocity differences are small (and thus there is no decoupling of
wind components).
On the other hand for stars for which the predicted mass-loss rates are
systematically too high {either} the {instability connected with the} decoupling of wind components may
occur {or} the frictional heating {increases wind temperature}.
Note however that {only for one star} we have obtained the possibility of
decoupling close to the stellar surface for velocities lower than the escape
velocity. {(Note that in order to obtain fall back of the wind material and
consequently lower wind mass-loss rate it is necessary to the decoupling
occur below the point where wind velocity is equal to the escape velocity.)}
{With this exception,} {presented} models are, to our knowledge, close to the star stable for perturbations of
multicomponent flow. This means that we are not able without any further
assumptions to explain \zm{very} low {observed} mass-loss rate{s} of {most} studied stars
{with $\dmdt\lesssim10^{-7}\,\smrok$} {(although in the outer wind regions of
these stars the multicomponent effects are important)}. However, there are several possibilities
how to (from the theoretical point of view) obtain decoupling of wind components
for  the velocities lower than the escape velocity also for other stars (like
slightly different metallicity, overestimation of theoretical mass-loss
rates\zm{, neglect of some heavier elements} or
improved treatment of the multicomponent flow).
{Another way how to explain low wind mass-loss rates may be connected with}
the thin-wind {effect} (Puls et al. \citeyear{puspo}, Owocki \& Puls
\citeyear{owpu}).


The decoupling discussed here slightly differs from that studied so far (cf.
Springmann \& Pauldrach \citeyear{treni}, KKII). 
{I}n the previous studies it was usually assumed that metals
and passive component \zm{may} decouple completely. {W}e present here
a more detailed and more realistic picture of the decoupling of wind components.
{W}e showed that some elements
which have very low abundance, however which significantly contribute to the
radiative force, may decouple \zm{(or cause the instability connected with the
decoupling)} {individually} from the mean flow, {that}  is
basically not decoupled. {Moreover,} the decoupling may
occur for higher mass-loss rates than it was previously assumed. {We present
an approximate formula for the test of importance of discussed decoupling of
individual heavier elements.}

{For our comparison we used mass-loss rates derived from the observation
assuming smooth winds. If the O star winds are clumped, as is indicated by
observations of e.g.~B03 or Martins et al. (\citeyear{okali}), then the  mass-loss
rates of O stars are likely lower than predicted by current hot
star wind theory and we obtain a disagreement between observations and theory
for all stars. On the other hand a better treatment of the opacity sources in
the UV domain with account of the line overlaps \zm{and comoving-frame radiative
transport} may help to overcome this
potential disagreement.}


We have studied the dependence of wind parameters on metallicity. It is
well-known that the mass-loss rate increases with {increasing} metallicity. We have derived
that the mass-loss  rate scales with metallicity as $\dmdt\sim Z^{0.67}$. The
terminal velocity of individual stars also varies with metallicity mainly due to
the {sensitivity of the radiative force on detailed wind state in outer wind
regions}. {However, o}n the average the terminal
velocity varies with metallicity only slightly (for studied values of
metallicity) {as} $v_\infty\sim Z^{0.06}$. As a consequence, the ratio of wind terminal
velocity to the surface escape velocity $v_\infty/v_\text{esc}\sim2.3$ is only
slightly lower than for Galactic stars.

\section*{Acknowledgements}

The author would like to thank Dr. Joachim Puls for valuable comments on the
manuscript and to Drs. Ji\v r\'\i\ Kub\'at and Zden\v ek Mikul\'a\v
sek for the discussion of this topic.
This research has made use of NASA's Astrophysics Data System and the
SIMBAD database, operated at CDS, Strasbourg, France.
This work was supported by grants GA \v{C}R 205/03/D020, 205/04/1267.


\begin{thebibliography}{}
{\bibitem[(1982)]{abpar} Abbott  D. C., 1982, ApJ, 259, 282}
\bibitem[(2003)]{bourak}  Bouret  J.-C., Lanz  T., Hillier  D. J.
 	et al., 2003, ApJ, 595, 1182 (B03)
\bibitem[(1996)]{zel6} Bautista  M. A., 1996, A\&AS, 119, 105
\bibitem[(1997)]{zel5} Bautista  M. A., \& Pradhan  A. K., 1997, A\&AS,
	126, 365
\bibitem[(1969)]{burgers} Burgers  J. M., Flow equations for composite
              gases, Academic Press, New York 1969
\bibitem[(1993)]{bumez} Butler  K., Mendoza  C., \& Zeippen  C. J., 1993,
	J. Phys. B., 26, 4409
{\bibitem[(1974)]{cassob} Castor  J. I., 1974, MNRAS 169, 279}
\bibitem[(1975)]{cak} Castor  J. I., Abbott  D. C., \& Klein  R. I., 1975,
	ApJ, 195, 157 (CAK)
\bibitem[(1993)]{uhlak} Charbonnel  C., Meynet  G., Maeder  A., Schaller  G.,
	\& Schaerer  D., 1993, A\&AS, 101, 415
\bibitem[(1999)]{zel3} Chen  G. X., \& Pradhan  A. K., 1999, A\&AS, 136,
	395
\bibitem[(2002)]{studeny} Crowther  P. A., Hillier  D. J., Evans  C. J.,
 	et al., 2002, ApJ, 579, 774
\bibitem[(2004)]{eva} Evans  C. J., Lennon  D. J., Trundle  C., Heap  S. R.,
	\& Lindler  D. J.,  2004, ApJ, 607, 451
{
\bibitem[(1997)]{felpulpal} Feldmeier  A., Puls  J., \& Pauldrach,
	A. W. A., 1997, A\&A, 322, 878}
{\bibitem[(2005)]{graham} Gr\"afener G., \& Hamann W.-R.,
        2005, A\&A, 432, 633}
\bibitem[(1992)]{hekuku} Herrero  A., Kudritzki  R. P., Vilchez  J. M.,
        et al., 1992, A\&A, 261, 209
\bibitem[(1991)]{hemalhut} Heydari-Malayeri  M., \& Hutsem\'ekers  D., 1991, 
	A\&A, 243, 401
\bibitem[(1998)]{hilmi} Hillier  D. J., \&  Miller  D. L., 1998, ApJ, 496,
	407
\bibitem[(1988)]{tlusty} Hubeny  I., 1988, Comput. Phys. Commun., 52, 103
\bibitem[(1992)]{hublaj} Hubeny  I., \& Lanz  T., 1992, A\&A, 262, 501
\bibitem[(1995)]{hublad} Hubeny  I., \& Lanz  T., 1995, ApJ, 439, 875
\bibitem[(1993)]{zel0} Hummer  D. G., Berrington  K. A., Eissner  W.,
	et al., 1993, A\&A, 279, 298
\bibitem[(2003)]{hubtueb} Hubeny  I., in Stellar Atmosphere
       Modelling  I. Hubeny  D.  Mihalas \& K. Werner eds., ASP Conf.
       Ser., Vol. 288, 17
\bibitem[(2001)]{kkii} Krti\v{c}ka  J., \& Kub\' at  J., 2001, A\&A, 377,
	175 (KKII)
\bibitem[(2002)]{kkiii} Krti\v{c}ka  J., \& Kub\' at  J., 2002, A\&A, 388,
	531
\bibitem[(2004)]{nltei} Krti\v cka  J., \& Kub\'at  J., 2004, A\&A, 417, 1003
{(KK1)}
\bibitem[(2004)]{velebobr} Krti\v cka  J. \& Kub\'at  J., 2005, in The A-Star
	Puzzle, IAU Symposium No. 224  J. Zverko  W.W. Weiss  J.
	\v{Z}i\v{z}\v{n}ovsk\'y \& S.J. Adelman, eds., {23}
\bibitem[(2003)]{gla} Krti\v{c}ka  J., Owocki  S. P., Kub\' at  J.,
	Galloway  R. K., \& Brown  J. C., 2003, A\&A, 402, 713
{
\bibitem[(1993)]{dis} Kub\'at  J., 1993, PhD. thesis, Astronomick\'y
	\'ustav AV \v{C}R, Ond\v{r}ejov}
\bibitem[(2003)]{kub} Kub\'at  J., 2003, in Modelling of Stellar
	Atmospheres, IAU Symp. 210  N. E. Piskunov  W. W. Weiss \& D. F.
	Gray eds., ASP Conf. Ser., {A8}
\bibitem[(1999)]{kpp} Kub\'at  J., Puls  J., \& Pauldrach  A. W. A.,
	1999, A\&A, 341, 587
{\bibitem[(1987)]{kupapu} Kudritzki R. P., Pauldrach A. W. A., \& Puls J., 1987,
	A\&A, 173, 293}
{\bibitem[(1989)]{kustar} Kudritzki R. P., Pauldrach A. W. A., Puls J., \&
	Abbott D.C., 1989, A\&A, 219, 205}
\bibitem[(2000)]{kupul} Kudritzki  R. P., \& Puls  J., 2000, ARA\&A, 38,
        613
{\bibitem[(2002)]{kudmet} Kudritzki  R. P., 2002, ApJ, 577, 389}
\bibitem[(1999)]{vald2} Kupka F., Piskunov  N. E., Ryabchikova, T .A.,
	Stempels  H. C., \& Weiss  W. W., 1999, A\&AS,  138, 119 
\bibitem[(1995)]{lsl} Lamers H. J. G. L. M., Snow T. P., \& Lindholm D.
	M., 1995, ApJ, 455, 269 
\bibitem[(1996)]{hmotak} Lanz  T., de Koter  A., Hubeny  I., \& Heap  S.
	R., 1996 ApJ, 465, 359
\bibitem[(2003)]{lahub} Lanz  T., \& Hubeny  I. 2003, ApJS, 146, 417
\bibitem[(1989)]{top1} Luo  D., \& Pradhan  A. K., 1989  J. Phys. B, 22,
	3377 
\bibitem[(2004)]{martin} Martins  F., Schaerer  D., Hillier  D. J.,
	\& Heydari-Malayeri  M., 2004, A\&A, 420, 1087 {(Mr04)}
\bibitem[(2005)]{okali} Martins  F., Schaerer  D., Hillier  D. J., 2005,
	A\&A, {436, 1049}
\bibitem[(2004)]{maso} Massey  P., Bresolin  F., Kudritzki  R. P., Puls  J.,
	\& Pauldrach  A. W. A., 2004, ApJ, 608, 1001 (M04)
{
\bibitem[(1974)]{sphermod} Mihalas  D., \& Hummer  D. G., 1974, ApJS 28,
	343}
\bibitem[(1969)]{mik} Mikul\'a\v sek  Z., 1969, BAICz, 20, 215
\bibitem[(1996)]{zel2} Nahar  S. N., \& Pradhan  A. K., 1996, A\&AS, 119,
	509
\bibitem[(1993)]{napra} Nahar  S. N., \& Pradhan  A. K., 1993  J. Phys. B,
	26, 1109
\bibitem[(1974)]{ng} Ng  K. C.  J. Chem. Phys., 61, 2680
\bibitem[(1999)]{owpu} Owocki  S. P., \& Puls  J., 1999, ApJ, 510, 355
\bibitem[(2002)]{op} Owocki  S. P., \& Puls  J., 2002, ApJ, 568, 965
\bibitem[(1987)]{pasam} Pauldrach  A. W. A., 1987, A\&A, 183, 295
\bibitem[(2001)]{pahole} Pauldrach  A. W. A., Hoffmann  T. L., \& Lennon,
	M., 2001 A\&A, 375, 161
\bibitem[(1990)]{pavyvoj} {Pauldrach  A. W. A., Kudritzki  R. P., Puls,
	J., \& Butler  K., 1990, A\&A, 228, 125}
\bibitem[(1995)]{vald1} Piskunov  N. E., Kupka F., Ryabchikova  T. A.,
	Weiss, \& W. W., Jeffery  C. S., 1995, A\&AS, 112, 525 
{
\bibitem[(1999)]{reac} Porter  J. M., \& Skouza  B. A., 1999, A\&A, 344,
	205}
\bibitem[(1998)]{rampa} Prinja  R. K., \& Crowther  P. A., 1998, MNRAS, 300, 828
\bibitem[(1996)]{pulmoc} Puls  J., Kudritzki  R.-P., Herrero  A.,
	et al., 1996, A\&A, 305, 171 (P96)
{
\bibitem[(2000)]{pusle} Puls  J., Springmann  U., \& Lennon  M., 2000,
	A\&AS, 141, 23}
\bibitem[(1998)]{puspo} Puls  J., Springmann  U., \& Owocki  S. P., 1998, in
Cyclical Variability in Stellar Winds  L. Kaper and A. W.
Fullerton eds.,  Springer-Verlag, 389
{
\bibitem[(2002)]{runow} Runacres  M. C, \& Owocki  S. P., 2002, A\&A, 381,
	1015}
\bibitem[(1992)]{rybashumremali} Rybicki  G. B. \& Hummer  D. G., 1992, A\&A, 262,
	209
\bibitem[(1992)]{savej} Sawey  P. M. J., \& Berrington  K. A., 1992,
	J. Phys. B, 25, 1451
\bibitem[(1987)]{top} Seaton  M. J., 1987  J. Phys. B, 20, 6363
\bibitem[(1992)]{topt} Seaton  M. J., Zeippen  C. J., Tully  J. A.,
	et al., 1992, Rev. Mexicana Astron. Astrofis., 23, 19
{
\bibitem[(1947)]{sobolevprvni} Sobolev  V. V. 1947, {\em Dvizhushchiesia
        obolochki zvedz}, Leningr. Gos. Univ., Leningrad
}
\bibitem[(1992)]{treni} Springmann  U. W. E., \& Pauldrach  A. W. A., 1992,
              A\&A, 262, 515 
\bibitem[(1999)]{ven} Venn  K. A., 1999, ApJ, 518, 405
\bibitem[(1999)]{vikolabis} Vink  J. S., de Koter  A., \& Lamers,
	H. J. G. L. M., 1999, A\&A, 350, 181
\bibitem[(2000)]{vikola} Vink  J. S., de Koter  A., \& Lamers,
	H. J. G. L. M., 2000, A\&A, 362, 295
\bibitem[(2001)]{vikolamet} Vink  J. S., de Koter  A., \& Lamers,
	H. J. G. L. M., 2001, A\&A, 369, 574 (VKL)
{%
\bibitem[(2006)]{ufosap} Votruba V., Feldmeier A., Kub\'at J.,
        \& Nikutta R., 2006, in Active OB-Stars: Laboratories for Stellar
	         \& Circumstellar Physics, S. \v Stefl, S. Owocki and A. Okazaki
		 eds}
\bibitem[(1996)]{zel1} Zhang  H. L., 1996, A\&AS, 119, 523
\bibitem[(1997)]{zel4} Zhang  H. L., \&  Pradhan A. K., 1997, A\&AS, 126,
	373
\end{thebibliography}
\end{document}